\documentclass[epj,nopacs]{svjour}
\usepackage{latexsym}
\usepackage{graphics}
\usepackage{graphicx}
\usepackage{bm}
\usepackage{epsfig}
\usepackage{amsmath,amsfonts}
\usepackage{url}
\usepackage{epstopdf}
\usepackage{color}
\newcommand{\be}{\begin{equation}}
\newcommand{\ee}{\end{equation}}
\newcommand{\bea}{\begin{eqnarray}}
\newcommand{\eea}{\end{eqnarray}}
\newcommand{\bse}{\begin{subequations}}
\newcommand{\ese}{\end{subequations}}
\newcommand{\bi}{\begin{itemize}}
\newcommand{\nn}{\nonumber}

\begin{document}

\title{Quasi-cosmological Traversable Wormholes in $f(R)$ Gravity}
\author{Hanif Golchin\thanks{\emph{e-mail:} h.golchin@uk.ac.ir} \and Mohammad Reza Mehdizadeh\inst{*}
\thanks{\emph{e-mail:} mehdizadeh.mr@uk.ac.ir}}  
%
%
\institute{Faculty of Physics, Shahid Bahonar University of Kerman, PO Box 76175, Kerman, Iran}
\date{Received: date / Revised version: date}

\date{Received: date / Revised version: date}

\abstract{In this paper we study traversable wormholes in the context of $f(R)$ gravity. Exact solutions of traversable wormholes are found by imposing the nonconstant Ricci scalar. These solutions asymptotically match spherical, flat and hyperbolic FRW metric. By choosing some  static $f(R)$ gravity models, we verify the standard energy conditions for the asymptotically spherical, flat and hyperbolic wormhole solutions. Unlike the Einstein gravity, we find that in the context of $f(R)$ modified gravity, the asymptotically spherical, flat and hyperbolic wormhole solutions can respect the null energy condition (NEC) at the wormhole throat and near that. 
We find that in some static $f(R)$ models, asymptotically flat and  hyperbolic wormholes respect the weak energy condition (WEC) through the whole space. 
\PACS{{PACS-key}{discribing text of that key}}}

\maketitle

\section{Introduction}
Traversable wormholes are throatlike geometrical structures which connect two separate and distinct regions of
spacetimes and have no horizon or singularity. There are some hints in the paper of L. Flamm \cite{fl}, to such spatial geometry. In 1935, Einstein and Rosen
\cite{eros} firstly obtained the wormhole solutions as bridge
model of a particle . The
study of Lorentzian wormholes in the context of general relativity (GR)
was stimulated by the significant paper of Morris and
Thorne in 1988 \cite{mtor}. A fundamental ingredient in wormhole physics is the flaring-out condition of the throat, which in GR
entails the violation of the NEC.  Matter that violates the NEC is denoted exotic matter.

A very important challenge in wormhole
scenarios is the establishment of standard energy conditions. In this regard, various methods have been proposed
in the literature that deal with the issue of energy conditions
within wormhole settings.  Work along this line has been
done in  dynamical wormholes \cite{kar2} and thin-shell wormholes \cite{Mehdizadeh:2015dta}, where the supporting matter is
concentrated on the wormholes throat. In the context of
modified theories of gravity, the presence of higher-order
terms in curvature would allow for building thin-shell 
wormholes supported by ordinary matter \cite{thin1}. Recently, many people try to build and study wormhole
solutions within the framework of modified gravity for
instance, wormhole solutions in BransDicke theory \cite{bran1}, Born-Infeld theory \cite{brn1}, Einstein-Gauss-Bonnet theory \cite{gaus1}, Kaluza-Klein
gravity \cite{kul1} and scalar-tensor gravity \cite{scal1} and Einstein-Cartan theory \cite{Bronnikov:2015pha,Bronnikov:2016xvj,ecart1}.

In recent years, theories of modified gravity have been studied involving cosmological objects like black holes, gravastars, strange stars and wormholes. Such theories are able to explain accelerating expansion of the universe and also solve the dark matter problem\cite{accel1}. A well- known theory of modified gravity is f(R) gravity which modifies GR  by replacing the  gravitational action R by an arbitary function f(R), where R is the Ricci scalar. These models can be  demonstrated to cause acceleration \cite{Starobinsky:1980te,mod1}. It was found that higher-order curvature invariants can satisfy the energy conditions in f(R) gravity \cite{lobener}. The power-law $R^m$ gravity is also studied in \cite{deb}, the new
static wormholes in f(R) theory using the non-commutative geometry are built in \cite{rahm1} and the cosmological evolution of wormhole solutions is investigated in \cite{char}. Recently, the junction conditions in f(R) 
applied to the construction of thin-shell wormholes  \cite{ero2} and pure double layer bubbles \cite{ero1}. The static wormhole geometries have also been theorized within curvature-matter coupling theories such as f(R, T) \cite{aziz1}. 

Lorentzian wormhole solutions were also investigated in viable f(R) modified theories of
gravity \cite{pmarko}, which are  consistent with
observations of the solar system and cosmological evolution in \cite{solar}. These wormholes were found in specific f(R) models which are considered to reproduce realistic scenarios of cosmological
evolution based on the accelerated expansion of the universe.  it was shown that the WEC holds at the vicinity of the throat for certain ranges of the free parameters of the theory. These asymptotically flat solutions  analysed  with the simple choices of shape function which  WEC is satisfied at a specific point in space. 

A large class of static black hole solutions  with non constant Ricci scalar  and  traversable wormholes  with constant Ricci scalar  have also investigated in the background of $f(R)$ gravity \cite{saba}, \cite{Aclas}.  The existence of wormhole solutions in scalar-tensor and $f(R)$ gravity is studied in \cite{Bronnikov:2006pt,Bronnikov:2010tt} and it is shown that static wormhole solutions in these theories obtain when the effective gravitational constant is negative (which means the anti-gravitational property of the solutions) in some region of space. 

In this work, our aim is to obtain  the wormhole solutions  with non constant Ricci scalar in the context of viable f(R) gravity. 
This paper is organized as follows: After this introduction, we will present the basic field equations of f(R) 
gravity . In Sec. III, we will derive the properties of  wormholes solutions within the framework
of f(R) gravity and discuss about
the energy conditions. The final section is devoted to
our concluding remarks.
\section{f(R) gravity wormholes}
The action of f(R) modified theories of gravity is given by
\be \label{action}
S=\frac{1}{2\kappa}\int d^4x\sqrt{-g}\,f(R)+\int d^4x\sqrt{-g}\,{\cal L}_m(g_{\mu\nu},\psi)\,,
\ee
where $\kappa=8\pi G$ and ${\cal L}_m$ is the matter
Lagrangian density, in which the matter field $\psi$ is minimally coupled to the metric $g_{\mu\nu}$. Throughout this work for notational simplicity we consider $\kappa=1$. One can find the field equation by varying the action (\ref{action}) with respect to metric $g^{\mu \nu}$ as
\be \label{feq}
FR_{\mu\nu}-\frac12\,f\,g_{\mu\nu}-\nabla_\mu \nabla_\nu
F+g_{\mu\nu}\Box F=T^m_{\mu\nu} \,,
\ee
and the trace of the field equation
\be \label{tr}
FR-2f+3\,\Box F=T \,,
\ee
where $F= df/dR$. Substituting the trace equation in (\ref{feq}) and re-organizing the terms, one can write the field equation as \cite{Lobo:2009ip}
\be \label{feq2}
G_{\mu\nu}\equiv R_{\mu\nu}-\frac12\,R\,g_{\mu\nu} =T^c_{\mu\nu}+\tilde{T}^{\,m}_{\mu\nu}\,.
\ee
In the above, the two terms in the right hand side of the field equation are in the form
\bea
T^c_{\mu\nu}&=&\frac{1}{F}\left[\nabla_\mu \nabla_\nu F
-\frac14\,g_{\mu\nu}(RF+\Box F+T) \right]\,, \nn \\ \tilde{T}^{\,m}_{\mu\nu}&=& T^{\,m}_{\mu\nu}/F \,.
\eea
In this background, we are going to study the static and spherically symmetric traversable wormholes which is given by the following metric\footnote{The static and spherically symmetric wormhole geometry is given by 
\be
ds^2=-e^{2\Phi(r)}dt^2+\frac{dr^2}{1-b(r)/r}+r^2\,(d\theta^2 +\sin
^2{\theta} \, d\phi ^2)\,. \nn
\ee
The solution is traversable when $\Phi(r)$ is finite. In the following we set $\Phi(r)=0$.}
\be \label{wmetric}
ds^2=-dt^2+\frac{dr^2}{1-b(r)/r}+r^2\,(d\theta^2 +\sin
^2{\theta} \, d\phi ^2) \,.
\ee
The Ricci scalar for the above wormhole geometry is
\be \label{ric}
R=\frac{2\,b'(r)}{r^2}\,,
\ee
where `` $'$ '' denotes derivative with respect to the radial coordinate $r$.
As in \cite{Lobo:2009ip} we choose the energy-momentum tensor of anisotropic distribution of matter
\be
T_{\mu\nu}=(\rho+p_t)U_\mu \, U_\nu+p_t\,
g_{\mu\nu}+(p_r-p_t)\,\chi_\mu \chi_\nu \,,
\ee
where $U^{\mu}$ is the four-velocity and $\chi^\mu=\sqrt{1-b(r)/r}\,\delta^{\mu}_r$ is the unit spacelike vector in the radial direction. $\rho(r)$ is the energy density and $p_r(r),\,p_t(r)$ are radial and transverse pressure respectively, so the energy-momentum tensor takes to the form $T^{\mu}_{\,\,\,\,\,\nu}={\rm
diag}[-\rho,p_r,p_t,p_t]\,$. Thus the field equation (\ref{feq2}) leads to the following relationships
\bea \label{feq3}
\frac{b'}{r^2}&=&\frac{\rho}{F}+\frac{H}{F}\,, \nn \\
-\frac{b}{r^3}&=&\frac{p_r}{F}+\frac{1}{F}\bigg\{\big(1\!-\frac{b}{r}\big) \left[F''\! -F'\frac{b'r-b}{2r^2(1-b/r)}\right]\! -H\bigg\} , \nn \\
\frac{b\!-\!b'r}{2r^3}     &=&\frac{p_t}{F}+\frac{1}{F}\left[\big(1-\frac{b}{r}\big)
     \frac{F'}{r} -H\right] , 
\eea
where $H(r)=\frac14\,(FR+\Box F +T)$. By solving the above system, one finds the following expressions \cite{Lobo:2009ip}
\bea
\rho&=&\frac{Fb'}{r^2}\,, \label{ro}   \\
p_r&=&-\frac{bF}{r^3}+\frac{F'}{2r^2}(b'r-b)-F''\big(1-\frac{b}{r}\big) \,, \label{pr}  \\
p_t&=&-\frac{F'}{r}\big(1-\frac{b}{r}\big)+\frac{F}{2r^3}(b-b'r)\,. \label{pt}
\eea
 We are interested in studying inhomogeneous spacetimes which merge smoothly to the cosmological background. Therefore, we may consider the Ricci scalar of the wormhole geometry  as
\be \label{expand}
R=6\,c_1+\frac{6\,c_2}{r^n}\,,
\ee
where $c_1,\, c_2$ and $n$ are free parameters. In the following section we consider some choices for these free parameters and study the corresponding wormhole solutions in some different $f(R)$ backgrounds.

\section{Traversable wormhole solutions in $f(R)$ models}
We start this section by finding wormhole solutions with Ricci scalar in the form (\ref{expand}). Combining (\ref{ric}) and (\ref{expand}) one can find the shape function as
\be \label{br}
b(r)=r^3\left(c_1+\frac{3\,c_2}{3-n}\,r^{-n}\right)+c_3,
\ee
where $c_3$ is the integration constant. This wormhole solution by choosing $c_2=0$ is obtained in \cite{Aclas}. In the following  we set the integration constant to zero. The shape function $b(r)$ for a wormhole solution, should satisfy the following conditions:
\bea \label{123}
&&{\it i}) \quad\,\, b(r_0)=r_0\,,\nn\\
&&{\it ii}) \quad\,\, b'(r_0)<1\,,\nn\\
&&{\it iii}) \quad\,\, 1-\frac{b(r)}{r}>0\, .
\eea
Inserting condition {\it i}), one can find $c_2$ in terms of $r_0,\,c_1$ and $n$; so it is possible to rewrite the shape function as
\be \label{bf}
b(r)=\left[\left( -r_0^{n}\,c_1+r_0^{n-2} \right) {r}^{3-n}+{c_1\,r^3} \right]\,.
\ee
 
\begin{figure*}
\begin{picture}(0,0)(0,0)
\put(80,-7){\footnotesize Fig (1.a)}
\put(238,-7){\footnotesize Fig (1.b)}
\put(402,-7){\footnotesize Fig (1.c)}
\end{picture}
\includegraphics[height=52mm,width=54mm]{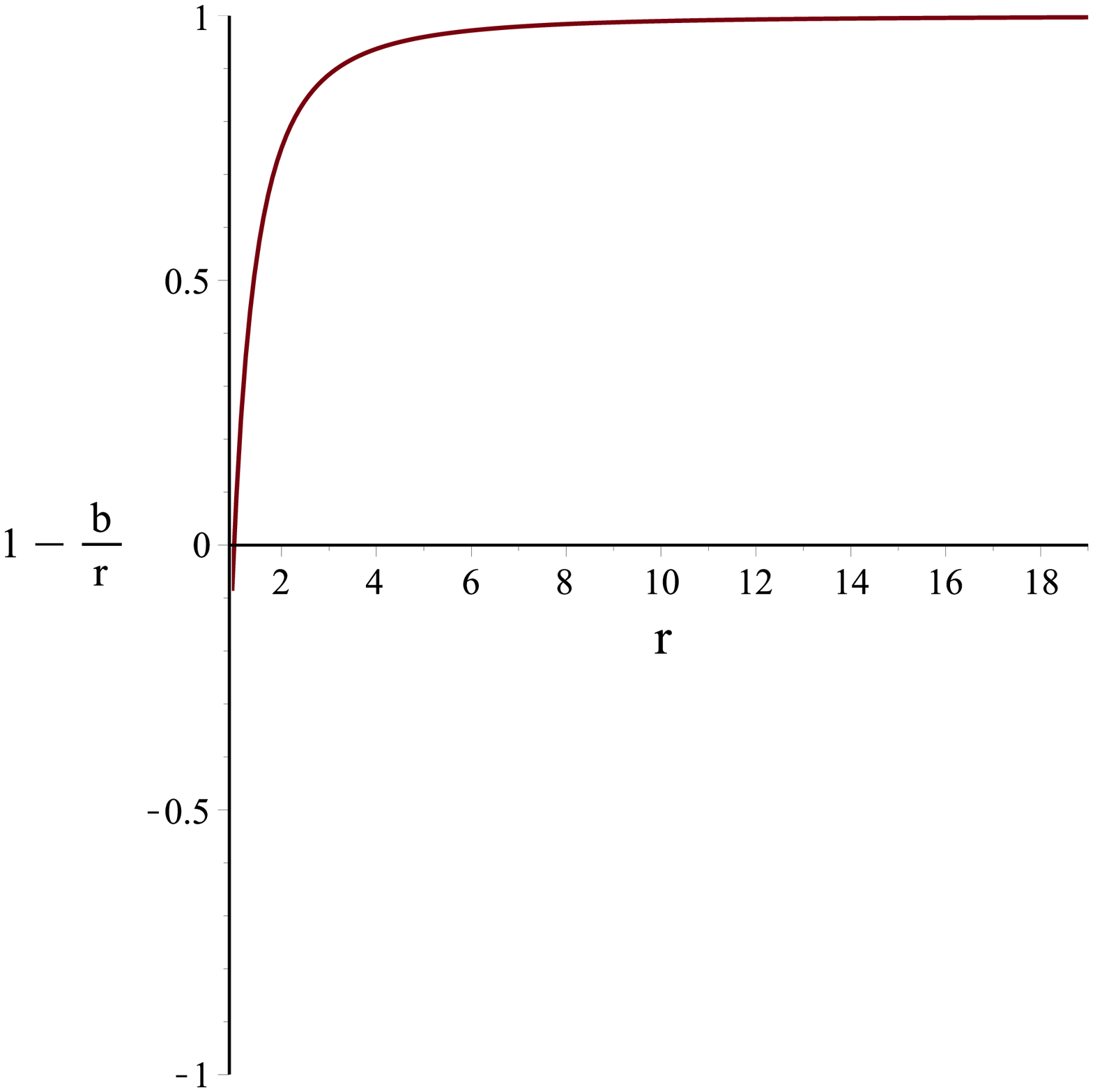} \,
\includegraphics[height=52mm,width=54mm]{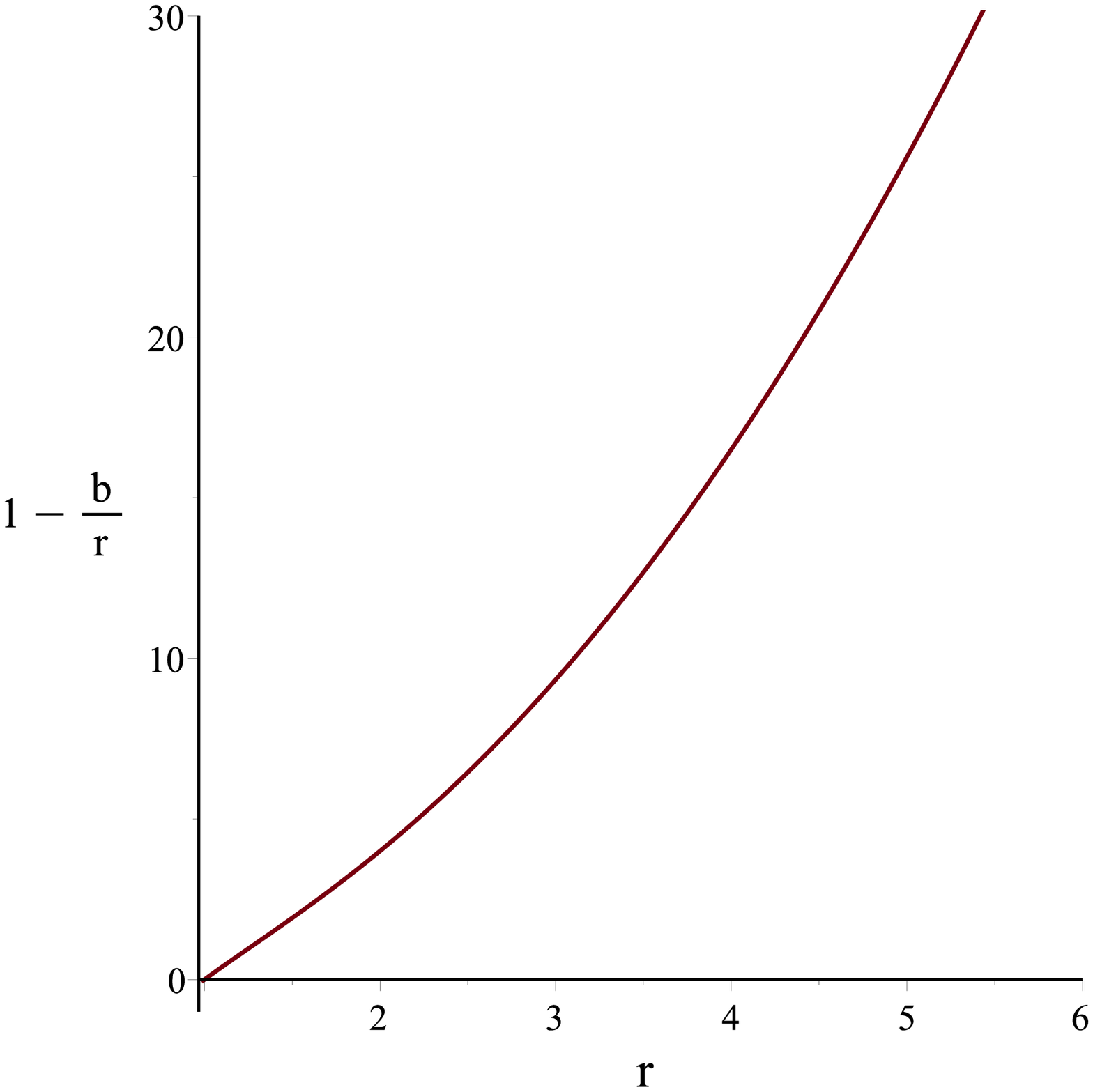} \, \includegraphics[height=52mm,width=52mm]{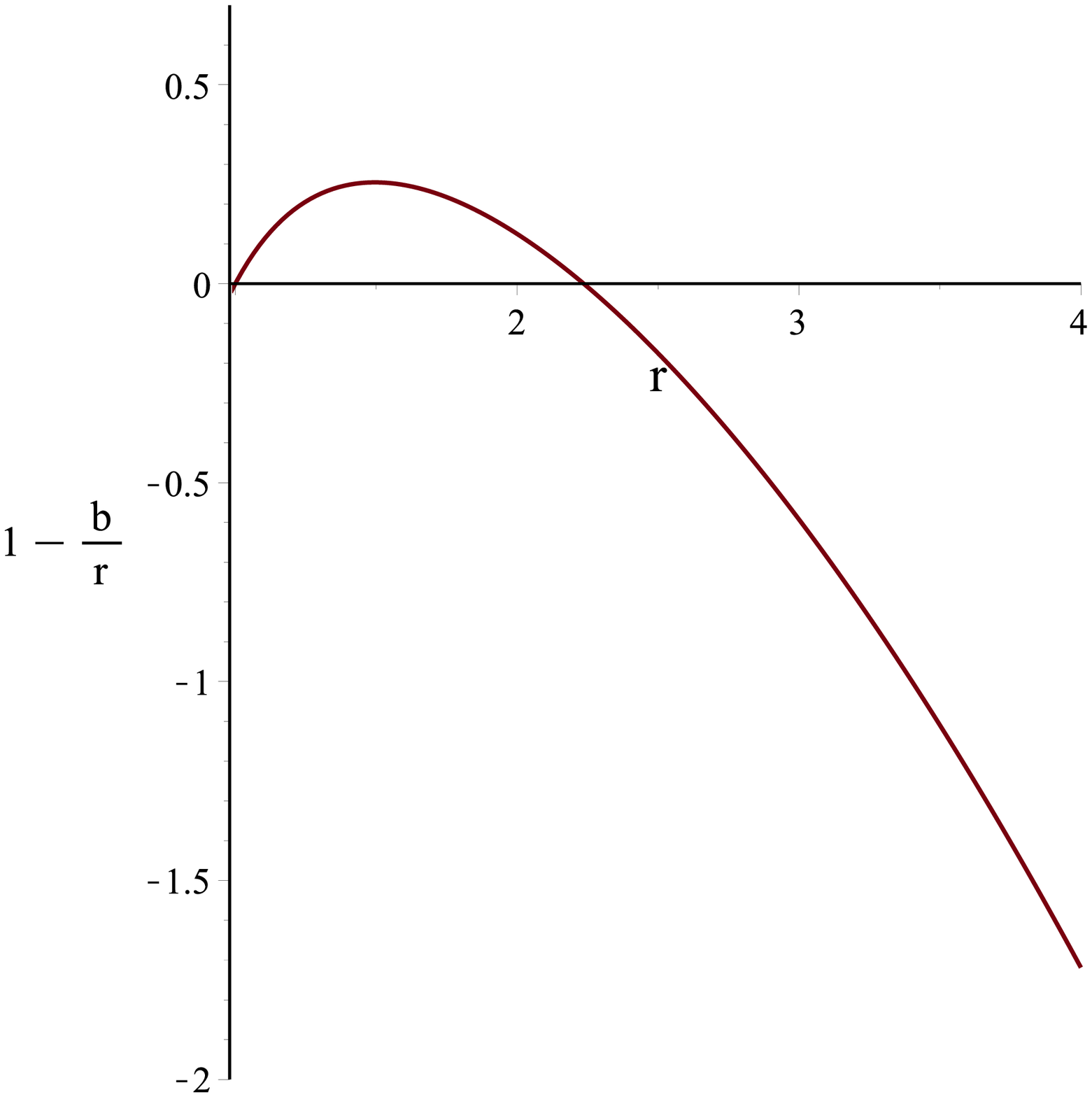}
\vspace*{0.8cm}       
\caption{Fig(1.a), Fig(1.b) and Fig(1.c) show the asymptotically flat ($c_1=0$), asymptotically hyperbolic ($c_1=-1$) and asymptotically spherical ($c_1=1$) wormhole solutions respectively. Wormhole throat is chosen at $r_0=1$, we also set $n=4$ in these figures.}\label{fig0}
\label{fig:2}       
\end{figure*}

We consider the wormhole solutions with three values $c_1=0,\pm 1$.
The shape function for these values are depicted in Fig \ref{fig0}. It is easy to check that in the case of $c_1=0,-1$ the condition {\it ii}) in (\ref{123}) is satisfied for $n>2$ and in the case of $c_1=1$ condition {\it ii}) is satisfied when $n>2.4$\,.

Setting the shape function into the form (\ref{bf}), one can write
\be \label{bff}
\frac{b(r)}{r}=c_1\,r^2+(r_0^{n-2}-c_1\,r_0^n)r^{2-n}\,,
\ee
note that due to the condition {\it ii}) we should insert $n>2$ in the above, so the second term in (\ref{bff}) falls down at large $r$. Remembering the wormhole metric (\ref{wmetric}), one deduces that the wormhole solutions with $c_1=-1,\,c_1=0$ and $c_1=1$ at large $r$ match the hyperbolic, flat and spherical FRW universe respectively,  so we call them asymptotically flat ($c_1=0$), asymptotically hyperbolic ($c_1=-1$) and asymptotically spherical ($c_1=1$) wormhole solutions.

In the framework of Einstein gravity, the traversable wormhole geometry (\ref{wmetric}), violates the weak energy condition WEC1: $\rho>0$\,, WEC2: $\rho+p_t>0$\,, WEC3: $\rho+p_r>0$\, and the null energy condition WEC2: $\rho+p_t>0$\,, WEC3: $\rho+p_r>0$\,. Substituting $\rho, p_r$ and $p_t$ given in (\ref{ro})-(\ref{pt}) and the shape function (\ref{bf}), one can rewrite the requirements of WEC as 
\bea \label{wec}
&&\!\!\!\!\!\!\!{\rm WEC1\!:} \left[3c_1+(n-3)r^{-n}(c_1\,r_0^n-r_0^{n-2})\right]F>0\,,\nn\\
&&\!\!\!\!\!\!\!{\rm WEC2\!:}\,\frac{1}{2r}\big[F\big(24\,c_1r\!+\!(4\!-\!n)(r_0^{n-2}\!\!-6c_1r_0^n)r^{1-n}\big)\nn \\
 &&\hspace{7mm}+2F'\!\left(c_1r^2\!\!-\!1\!+\!(r_0^{n-2}\!-6c_1r_0^n)r^{2-n}\right)\big]\!>0,\nn\\
&&\!\!\!\!\!\!\!{\rm WEC3\!:}\,F\left[12c_1 -(n-2) \left( r_0^{n-2}-6c_1r_0^{n} \right)r^{-n}\right]\nn \\
&& \hspace{7mm}+\frac12 F'\!\left[12c_1r+ (6 c_1r_0^{n}- r_0^{n-2} )  \left( n-2 \right) {r}^{1-n}\right]\nn\\
&&\hspace{7mm}+ F'' \! \left[6c_1r^2\!-\!1\!+\! (r_0^{n-2}\!-\!6\,r_0^{n}c_1 ) \right]\! {r}^{2-n}>0,
\eea
where $F=df(R)/dR$ and `` $'$ '' denotes derivative with respect to the radial coordinate $r$. 

The conditions of stability and absence of ghosts are discussed in \cite{Bronnikov:2006pt,Bronnikov:2010tt}. It is shown that in a scalar-tensor theory of gravity formulated in the Jordan frame with Lagrangian
\be
L=\frac12\left[f(\phi)R+h(\phi)g^{\mu \nu}\phi_{, \mu} \phi_{, \nu}-2U(\phi)\right]+L_m\,,
\ee
where $f$, $h$ and $U$ are arbitrary functions and $L_m$ is the matter Lagrangian, there is no static wormhole solution that satisfy the null energy conditions when $f(\phi)>0$ and $f(\phi)\,h(\phi)+\frac32\big(\frac{df}{d\phi}\big)^2>0$. In the case of $f(R)$ theories of modified gravity $h(\phi)=0$ and $f(\phi)=F=df(R)/dR$ \cite{Bronnikov:2010tt}, so the above nonexistence theorem holds for
\be \label{nogo}
F=\frac{df(R)}{dR}>0\,, \qquad \frac{d^2 f(R)}{dR^2}\ne 0\,.
\ee

In the following we consider the asymptotically flat ($c_1=0$), asymptotically hyperbolic ($c_1=-1$) and asymptotically  spherical ($c_1=1$) wormhole solutions in the background of $f(R)$ modified gravity. We are interested to the wormholes that respect  the requirements of WEC, in the other word, we look for the solutions which satisfy (\ref{wec}), the conditions {\it ii}), {\it iii}) in (\ref{123}) and violate the nonexistence theorem (\ref{nogo}) simultaneously\footnote{Note that by $d^2 f(R)/dR^2=0$ in (\ref{nogo}), the nonexistence theorem violates just in one or some points instead, the $F<0$ violates this theorem in a region at space. Hence in the following we search for $F<0$ to check the violation of this nonexistence theorem.}. Note that the regions where $df(R)/dR < 0$ are anti-gravitational, in the sense that the effective gravitational constant is negative \cite{Bronnikov:2006pt,Bronnikov:2010tt}.

\subsection{The case of $c_1=0$}
For a wormhole solution, as we mentioned above, the second term in (\ref{bff}) falls down at large $r$, therefore in the case of  $c_1=0$ the traversable wormhole solution (\ref{wmetric})  matches the flat FRW metric at large $r$. At this stage we consider such asymptotically flat wormholes in the background of some static $f(R)$ gravity models. As we will see, the $f(R)$ modified gravity wormhole solutions in many cases satisfy the null or weak energy condition.
 
\begin{figure*}[ht]
\begin{picture}(0,0)(0,0)
\put(124,-211){\footnotesize Fig (2.a)}
\put(355,-211){\footnotesize Fig (2.b)}
\end{picture}
\center
\includegraphics[height=67mm,width=75mm]{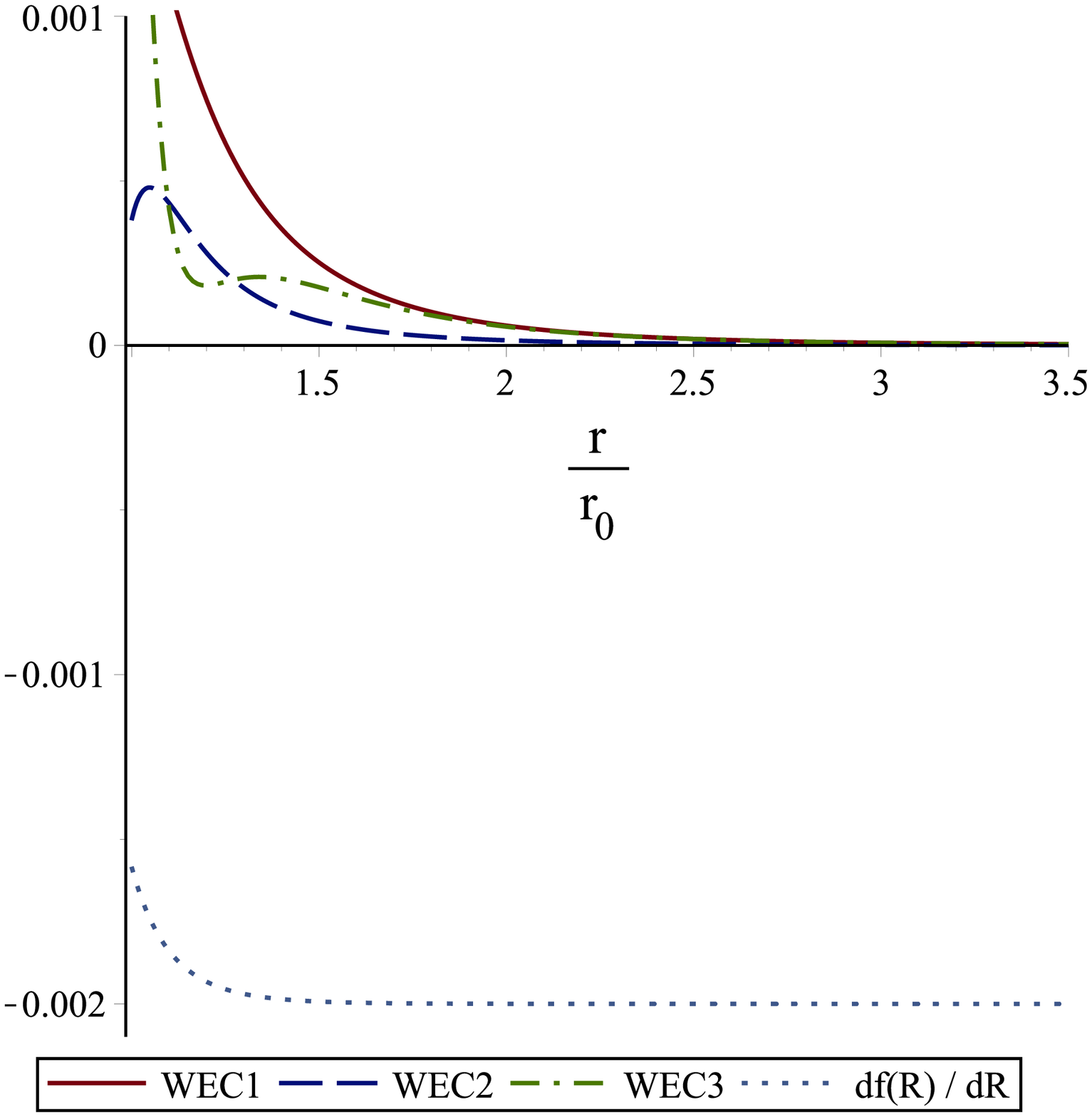} \quad \includegraphics[height=67mm,width=75mm]{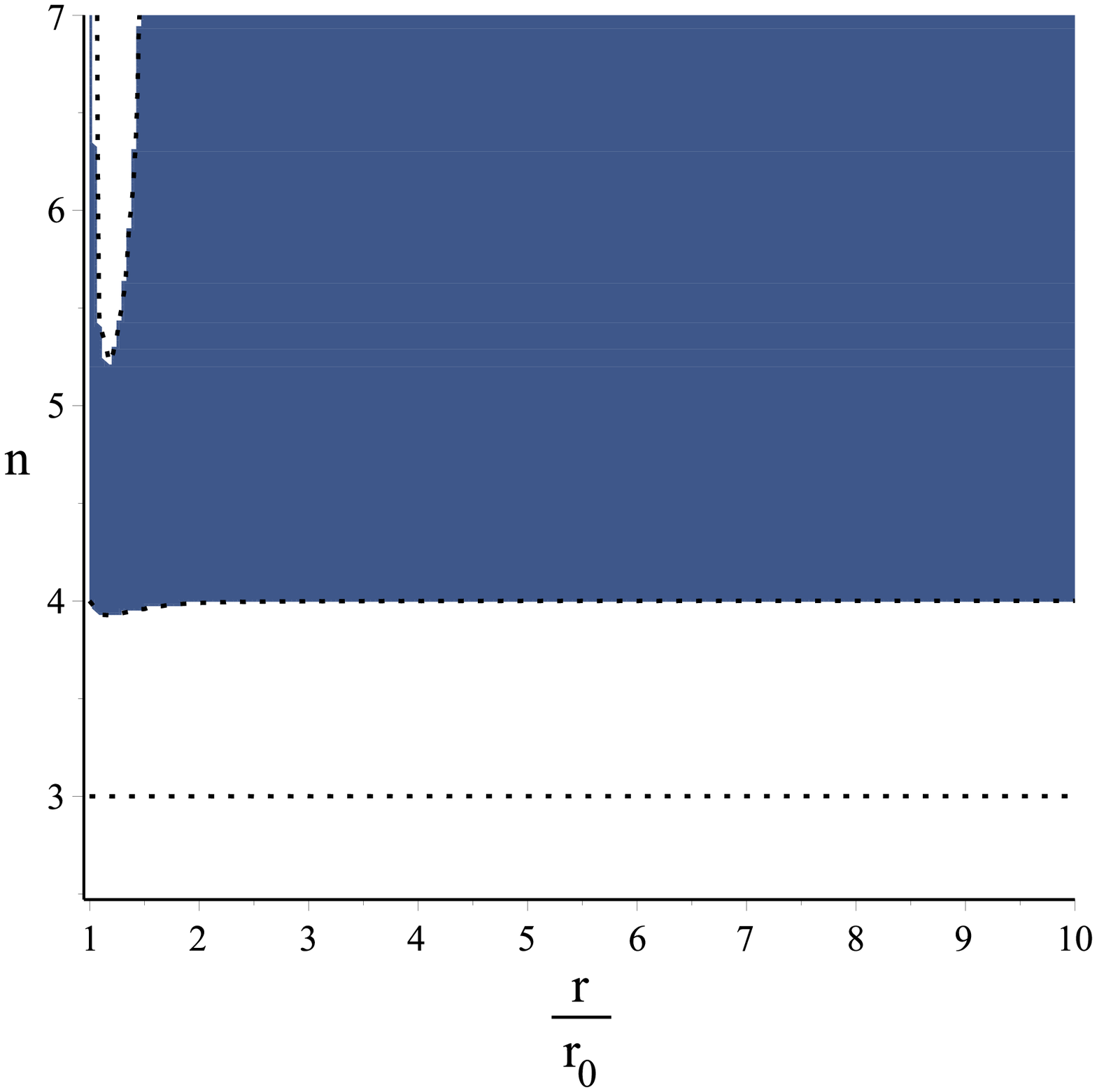}
\vspace*{8mm}
\caption{{\small $F=df(R)/dR$ and weak energy condition requiements for a wormhole solution of the Tsujikawa model with $n=5$ are depicted in Fig(2.a). It is obvious that $F<0$, so the condition of wormhole nonexistence is violated. It is also clear that WEC is respected at the throat $r_0$ and beyond it. In the blue region of Fig(2.b), $F<0$ and (\ref{wec}) are simultaneously satisfied, so this region shows traversable asymptotically flat wormholes of the Tsujikawa model, that respect the WEC. In these figures we set $\mu=1.02, R_*=0.1$ and $r_0=25$. }\label{figtsu}}
\end{figure*}

 \vspace{2mm}
{\bf 1)} The Tsujikawa model \cite{Tsujikawa:2007xu,DeFelice:2010aj,luca}\\
One of $f(R)$ models which is consistent with cosmological and local gravity conditions, proposed by Tsujikawa. In this viable model, the $f(R)$ considered as
\be \label{tsu}
f(R)=R-\mu R_* \tanh(\frac{R}{R_*})\,,
\ee
where $0.905<\mu<1$ and $R_*$ is a free small positive parameter \cite{Tsujikawa:2007xu}. For this model one can easily finds $F=1-\mu+\mu \tanh^2(R/R_*)$. It is obvious that $F$ takes positive values when $\mu<1$, this means that by choosing $\mu$ in the consistent range $0.905<\mu<1$, due to (\ref{nogo}), there is no static wormhole solution in this model that satisfy the  NEC (and so  WEC). However by taking $\mu$ at the neighborhood of this range, one can find wormholes that respect WEC. In figure \ref{figtsu}.a we have depicted $F$ in (\ref{nogo}) and the requirements of WEC (\ref{wec}) for a solution with $\mu=1.02$ and $n=5$. It is obvious in this figure that $F<0$\,, therefore the condition of nonexistence theorem (\ref{nogo}) is violated. The figure also shows that  WEC is respected at the wormhole throat $r_0$.

It is worth to mention that in this model, there is a 4-dimensional parameter space for the wormhole solutions which is made of $\mu, R_*, r_0$ and $n$. Among these parameters, $\mu, R_*$ are limited by the model \cite{Tsujikawa:2007xu}, we also measure the radial distance $r$, in the unit of wormhole throat $r_0$ (the WEC1,2,3 in figure \ref{figtsu}.a are depicted in term of $r/r_0$). Now we are going to find the parameter $n$ in a manner that wormhole solutions respect WEC through the space. In the other word we search for regions in $n-r/r_0$ diagram that (\ref{wec}) and conditions {\it ii}), {\it iii}) in (\ref{123}) are satisfied and $F$ in (\ref{nogo}) takes a negative value simultaneously.

Figure \ref{figtsu}.b shows $n-r/r_0$ diagram. The blue region in this figure corresponds to asymptotically flat traversable wormhole solutions in the background of (\ref{tsu}), which respect the WEC. It is obvious that by choosing $4<n<5.2$, the WEC is respected at the throat $r_0$ and through the whole space outside that. Note that unlike the Einstein gravity, considering $f(R)$ gravity model in the form (\ref{tsu}), it is possible to find traversable wormhole solutions that respect the WEC at the throat and beyond that.

\begin{figure*}[ht]
\begin{picture}(0,0)(0,0)
\put(125,-217){\footnotesize Fig (3.a)}
\put(362,-217){\footnotesize Fig (3.b)}
\end{picture}
\center
\includegraphics[height=70mm,width=80mm]{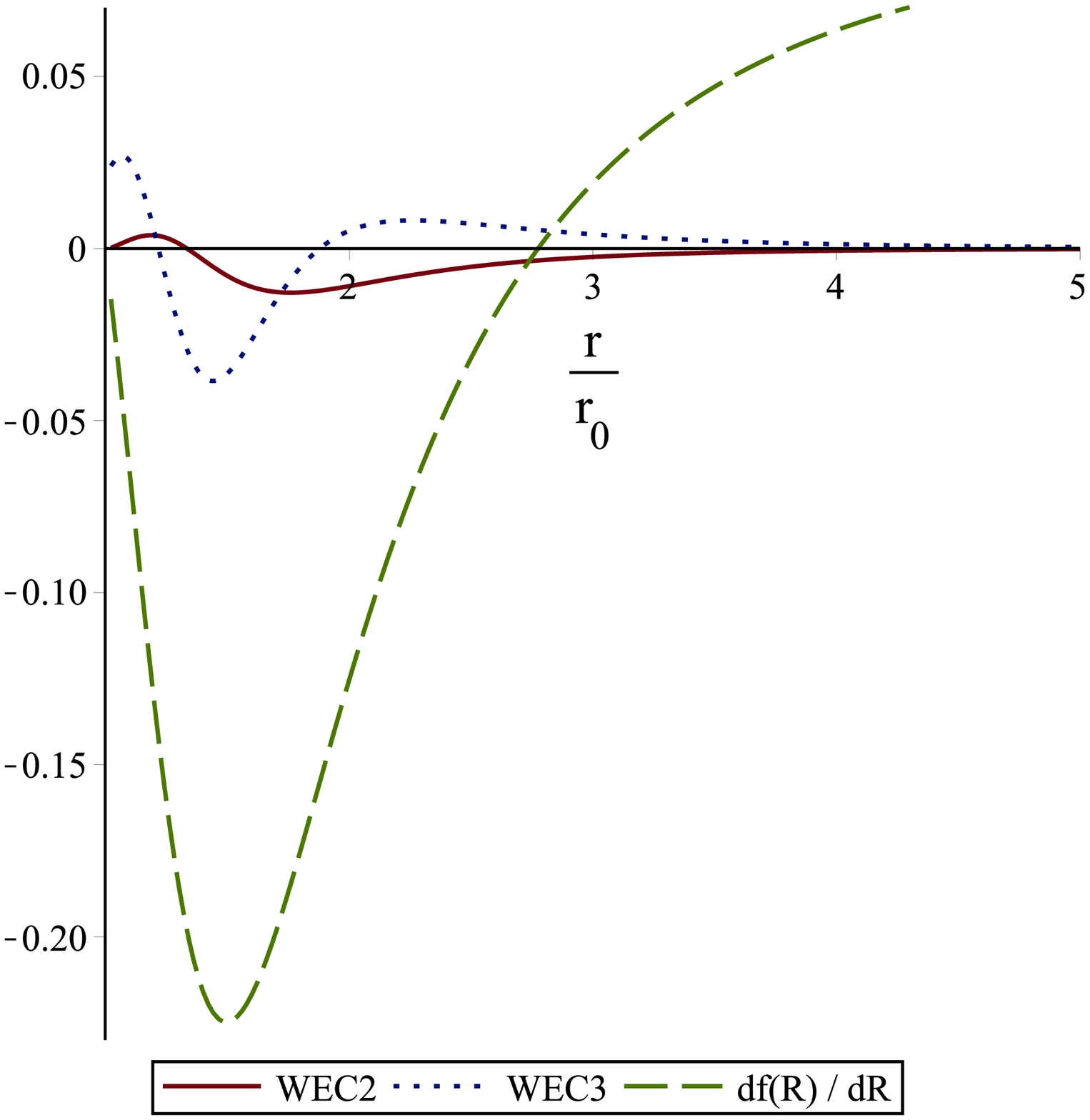} \quad \includegraphics[height=70mm,width=75mm]{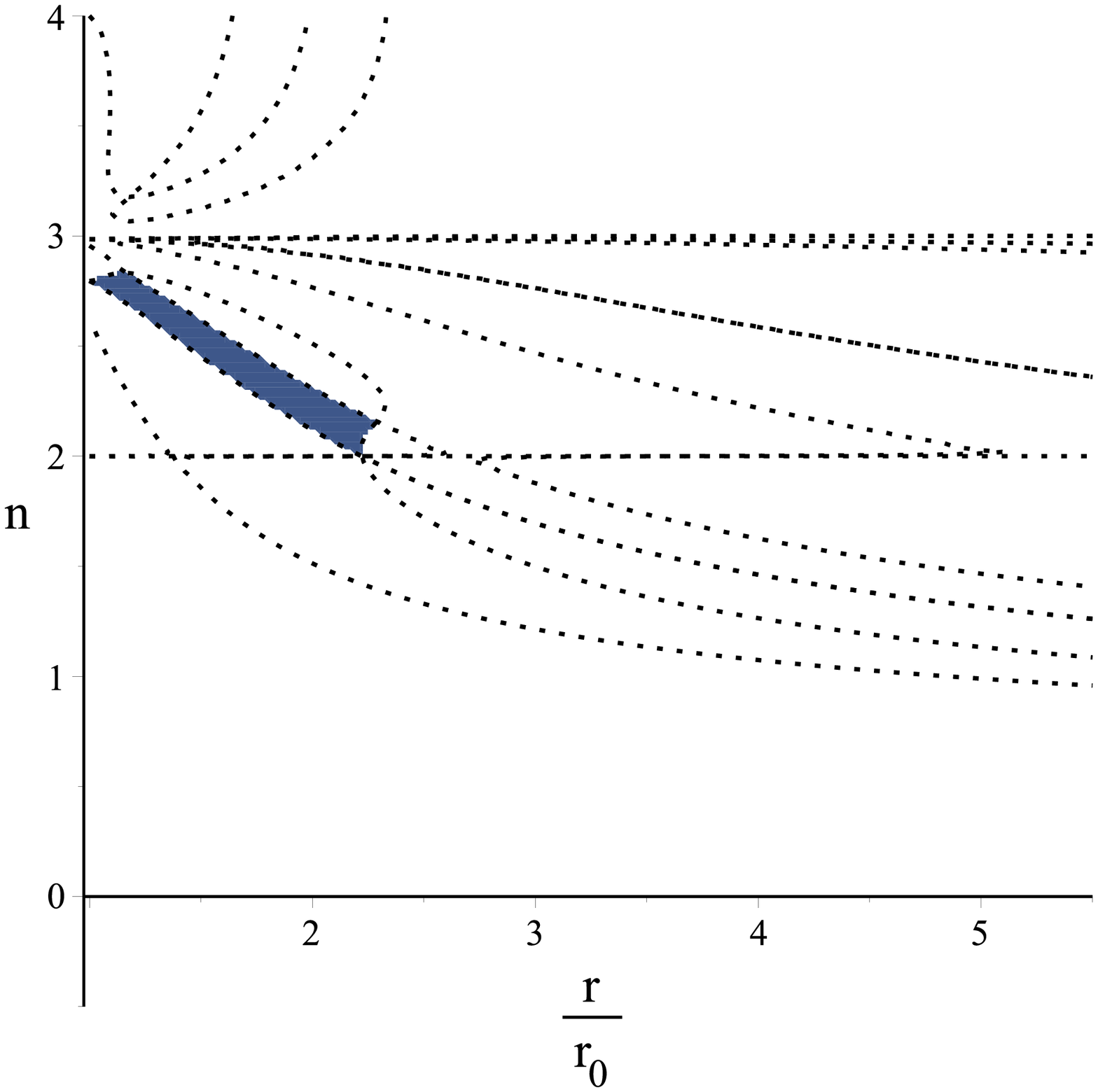}
\vspace{8mm}
\caption{{\small Setting $n=2.8$ in Fig(3.a), we can see that $df(R)/dR<0$ around the throat and the obtained wormhole solution of Hu-Sawicki model respect NEC at $r_0$.  The blue region in Fig(3.b) corresponds to traversable asymptotically flat wormholes of the Hu-Sawicki model, which respect NEC. In these figures we set $m=1, \mu=5\,, R_*=0.001$ and $r_0=15$. }\label{fighs}}
\end{figure*}

\vspace{3mm}
{\bf 2)} Hu-Sawicki model\\
The second $f(R)$ model which satisfies both cosmological and local gravity constraints proposed by Hu and Sawicki \cite{Hu:2007nk,luca}. This model is in the form 
\be \label{hs}
f(R)=R-\mu R_* \frac{(R/R_*)^{2m}}{(R/R_*)^{2m}+1}\,,
\ee
where $m, R_*$ and $\mu$ are positive parameters. It is shown in \cite{Tsujikawa:2007xu} that $R_*$ takes a small positive value and for $m=1$ one should insert $\mu \ge 8\sqrt{3}/9$. Similar to the previous case,  we are looking for the wormhole solutions of this $f(R)$ model which respect the NEC or WEC. In figure \ref{fighs}.a by setting $\mu=5$ and $n=2.8$, we have depicted $F$ in (\ref{nogo}) and WEC2,3 (\ref{wec}). It is clear from the figure that around the throat $r_0$, $F<0$ and WEC2,3 are positive, which means that this wormhole solution respect the null energy conditions.
 
Figure \ref{fighs}.b shows the $n-r/r_0$ diagram for this model. The blue zone in this figure is the region that $F<0$, ${\rm WEC2>0}$, ${\rm WEC3>0}$ and {\it ii}), {\it iii}) in (\ref{123}) are satisfied simultaneously. This means that traversable asymptotically flat wormhole solutions in this model respect the NEC around the throat $r_0$ when $2<n<3$ and by setting $n\sim 2.8$ the NEC is  respected at the throat.

\begin{figure*}[ht]
\center
\includegraphics[height=70mm,width=80mm]{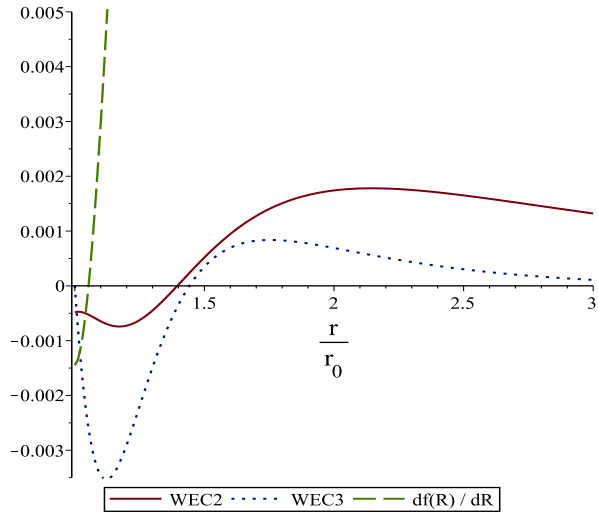} 
\caption{{\small It is clear that $df(R)/dR<0$ at the throat, but the wormhole solutions in Starobinsky model do not respect NEC at the throat $r_0$. In this figures we set $q=2, \lambda=0.98\,, R_*=0.01$ and $r_0=15$. }\label{figstar}}
\end{figure*}

\vspace{3mm}
{\bf 3)} The Starobinsky model \cite{Starobinsky:2007hu,Amendola:2006kh,luca}\\
Another viable modified $f(R)$ gravity model that we consider, has three free parameters $\lambda, R_*$ and $q$ as:
\be \label{staro}
f(R)=R+\lambda R_*\left[\big(1+\frac{R^2}{R_*^2}\big)^{-q}-1\right],
\ee
where $0.944<\lambda<0.966$ for $q=2$ according to \cite{Tsujikawa:2007xu} and $R_*$ takes small positive values. Similar to the previous case, we are looking for the wormhole solutions of this $f(R)$ model which respect the WEC (\ref{wec}).  One can check that by choosing $0.944<\lambda<0.966$, $F$ in (\ref{nogo}) takes positive values and there is no static wormhole solution in this model. However by taking $\lambda$ at the neighborhood of this range, it is possible to obtain negative values for $F$. In figure \ref{figstar}.a we draw $F$ and (\ref{wec}) by choosing $\lambda=0.98, n=2.5$. It is clear that by this choice, $F$ is negative at the throat however the null (and so weak) energy condition  is violated at the wormhole throat.

We also tried -by varying the parameter $n$- to find wormhole solutions that respect NEC, but there is no region in $n-r/r_0$ diagram that $F<0$ and  the NEC is satisfied simultaneously (the $n-r/r_0$ diagram is blank). In the other words, our results show that there is no asymptotically flat wormhole in this model that respect the NEC.

\begin{figure*}[ht]
\begin{picture}(0,0)(0,0)
\put(105,-7){\footnotesize Fig (5.a)}
\put(338,-7){\footnotesize Fig (5.b)}
\end{picture}
\includegraphics[height=70mm,width=79mm]{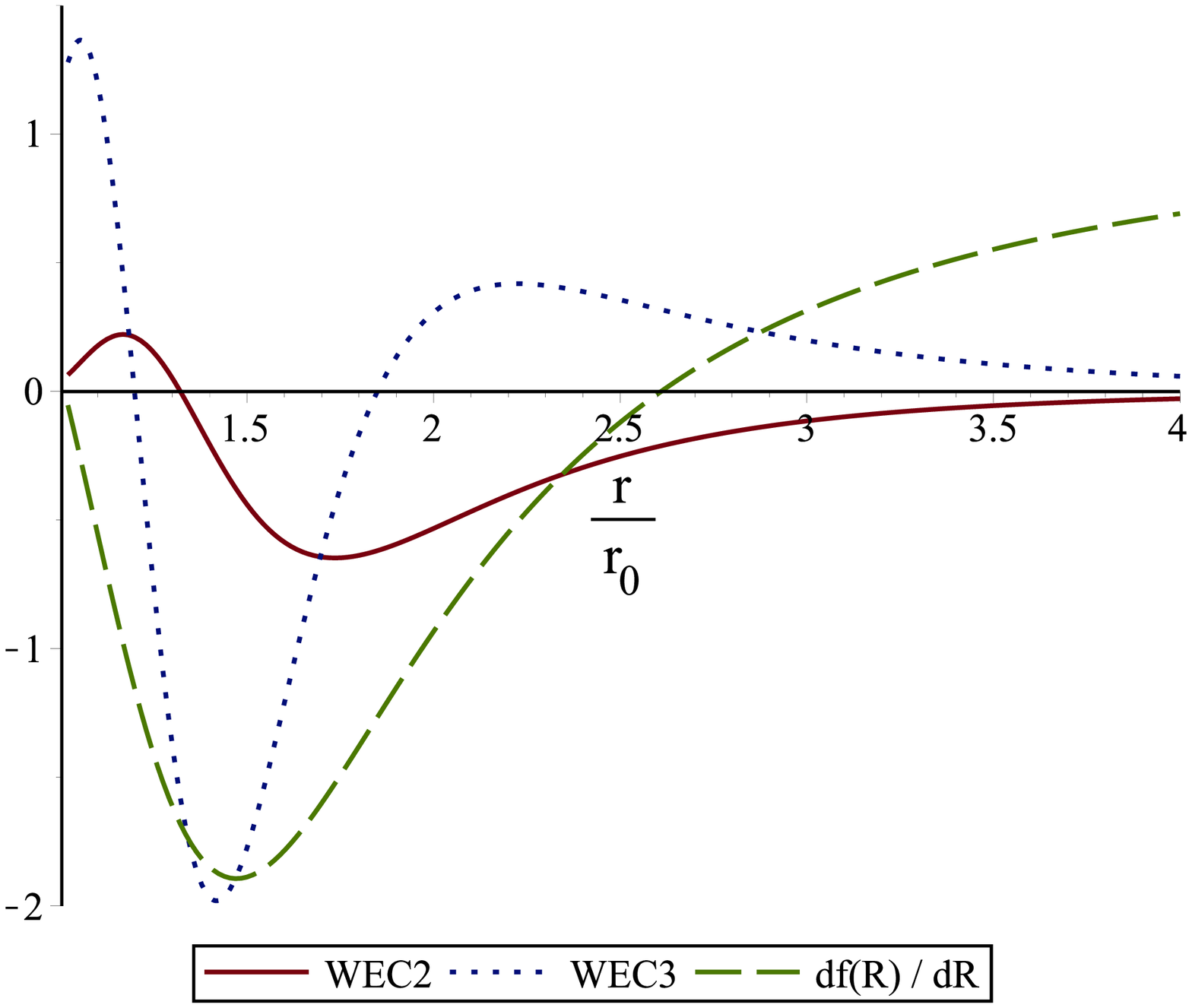} \quad \includegraphics[height=70mm,width=75mm]{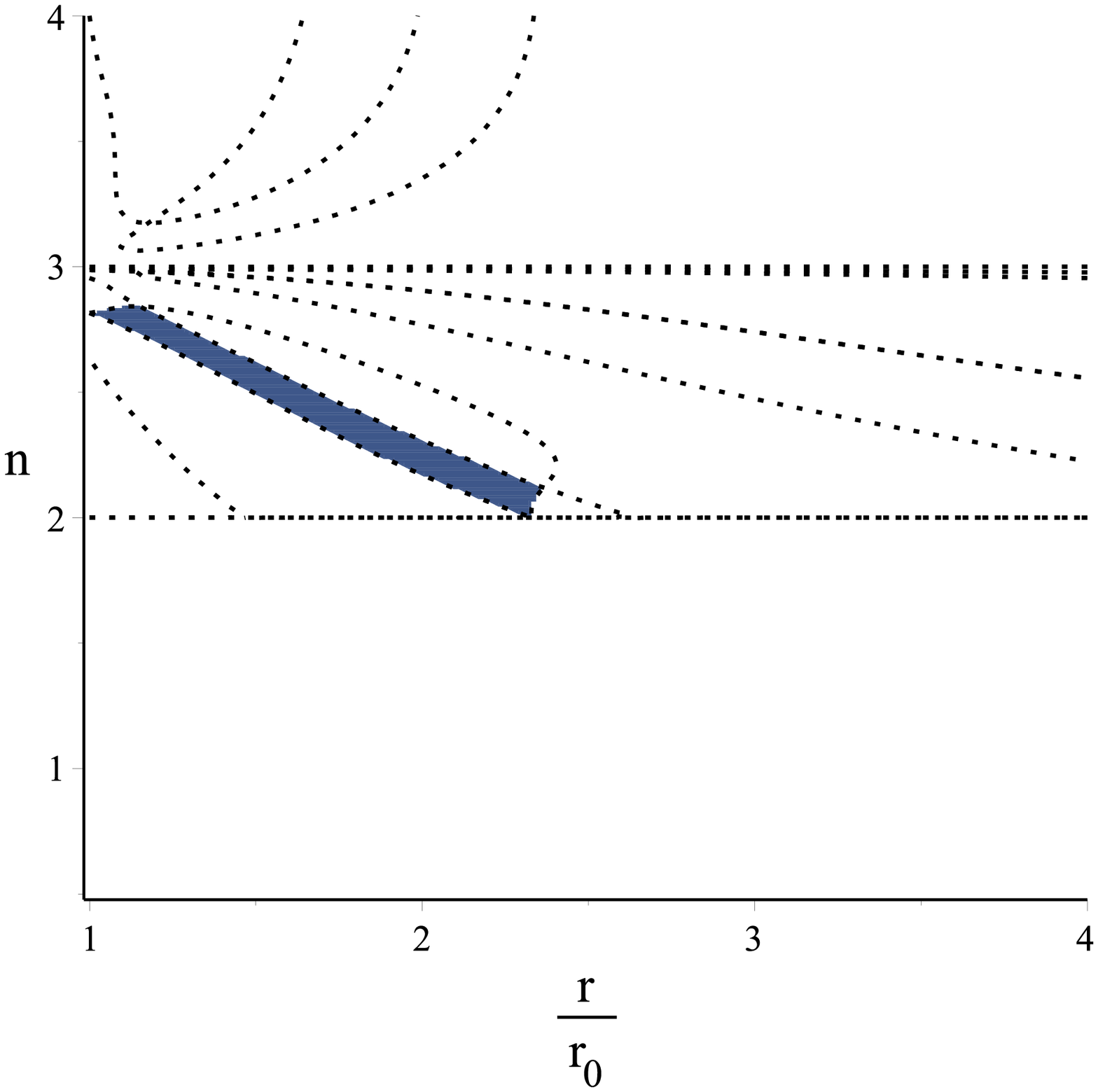}
\vspace{8mm}
\caption{{\small Fig(5.a) shows that $df(R)/dR<0$ and the NEC is satisfied at the throat for the wormhole solution of the Nojiri-Odintsov model with $n=2.8$.  According to Fig(5.b) asymptoticly flat traversable wormholes in this model, respect NEC  by choosing adequate values for $n$. In these figures we set $m=2,\, a=2,\, b=20,\, c=20$ and $r_0=1$.}\label{n-o}}
\end{figure*}

\vspace{3mm}
{\bf 4)} Nojiri-Odintsov model \cite{n-o}\\
Another viable $f(R)$ model suggests that
\be \label{no}
f(R)=R+\frac{R^m\big(aR^m-b\big)}{1+cR^m}\,,
\ee
where $m\geq2$ and $a, b$ and $c$ are free positive parameters which satisfy the condition $bc>>a$. Setting $m=2,\, a=2,\, b=20,\, c=20$ and $n=2.8$, we checked that $F<0$ and WEC2, WEC3 are respected around the wormhole throat $r_0$ (figure \ref{n-o}.a). Figure \ref{n-o}.b shows that the $f(R)$ model (\ref{no}) contains asymptoticly flat wormhole solutions with different values for the parameter $n$, that respect the null energy condition.

\begin{figure*}[ht]
\begin{picture}(0,0)(0,0)
\put(120,-217){\footnotesize Fig (6.a)}
\put(355,-217){\footnotesize Fig (6.b)}
\end{picture}
\center
\includegraphics[height=70mm,width=75mm]{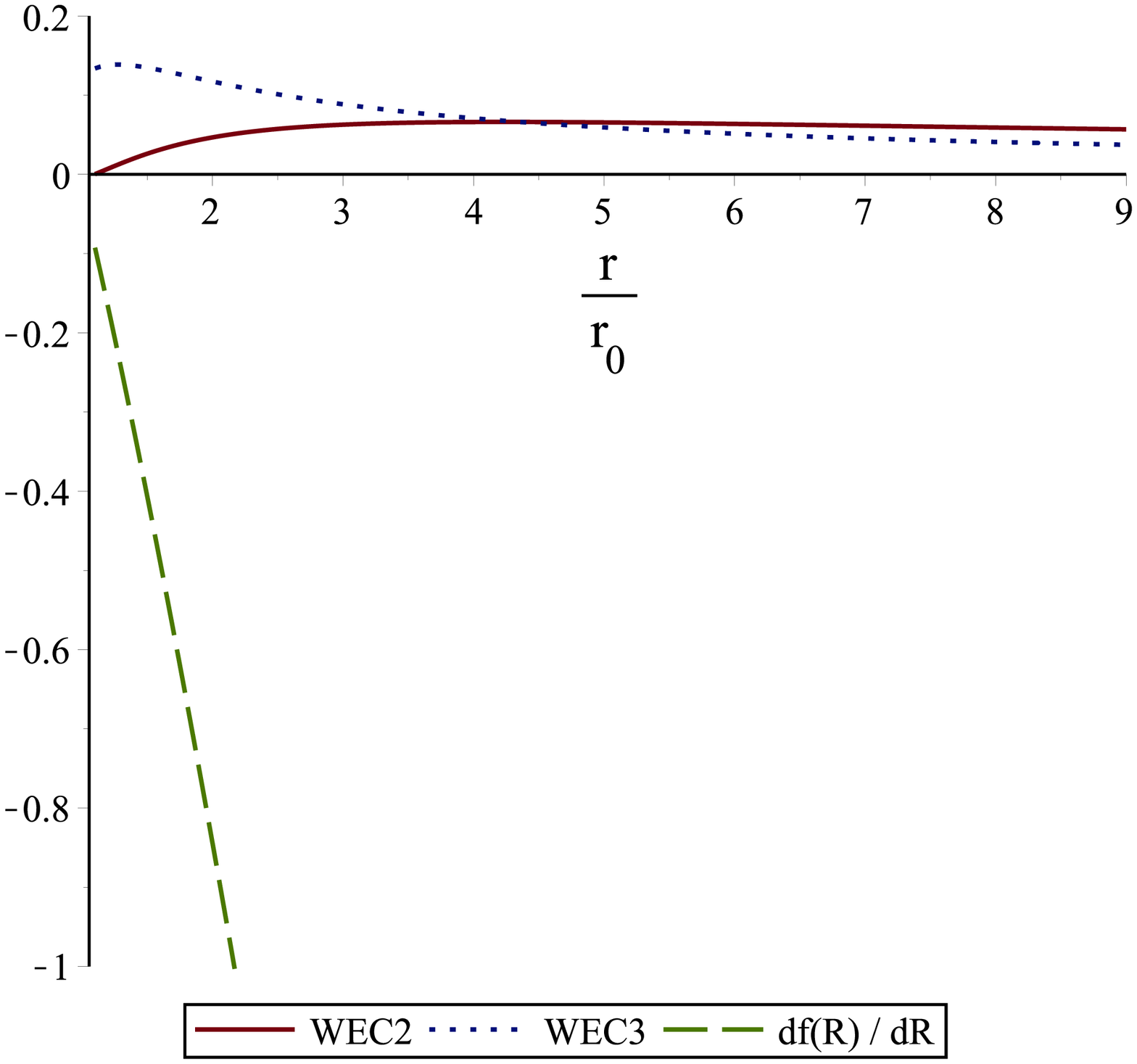} \quad \includegraphics[height=70mm,width=75mm]{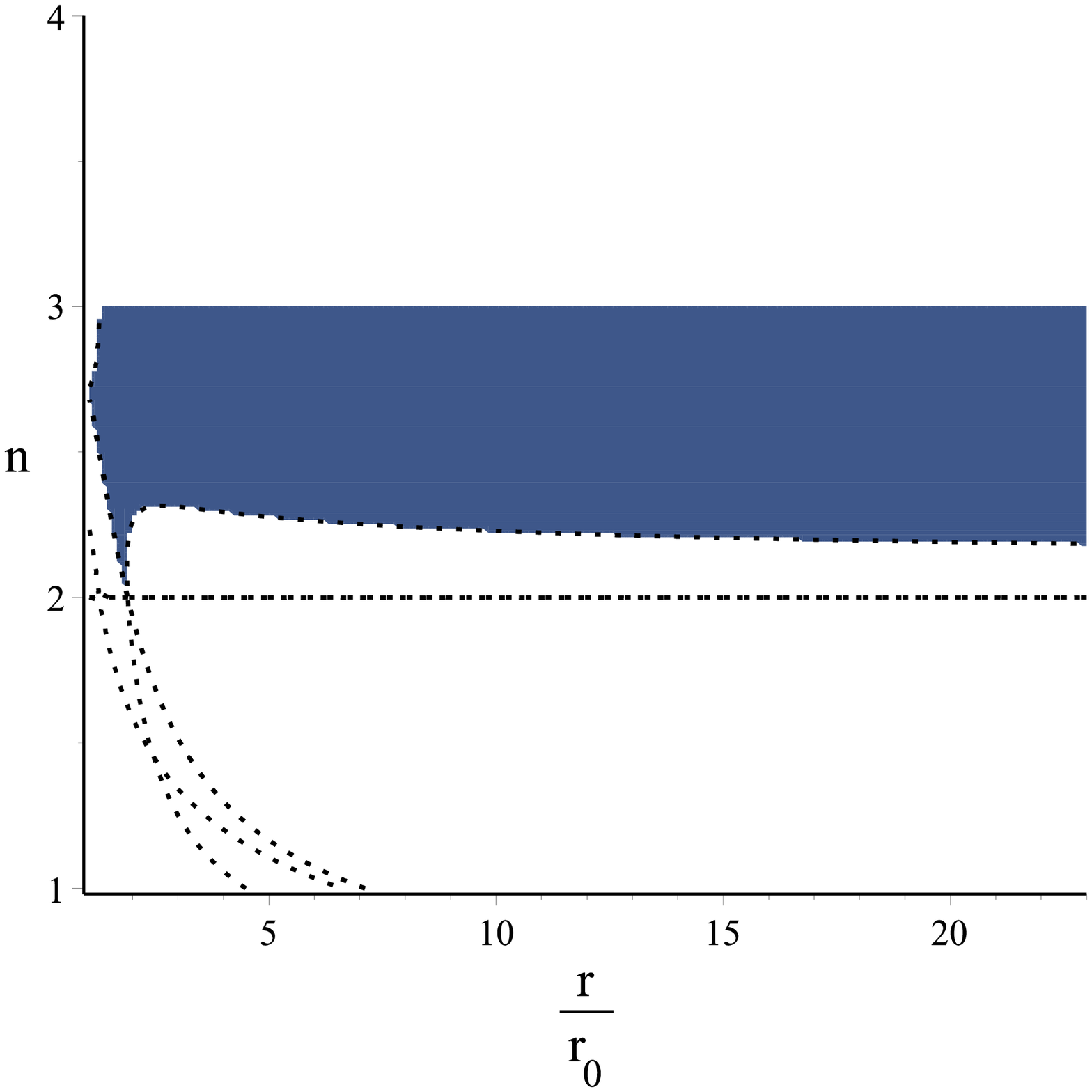}
\vspace{8mm}
\caption{{\small The NEC for a wormhole solution in the case of  Amendola-Gannouji-Polarski-Tsujikawa model is respected (Fig 6.a). In the blue part in Fig(6.b), traversable asymptoticly flat wormholes respect the NEC. In these figures we set $\mu=5, R_*=0.01, p=0.5$ and $r_0=3$. }\label{agp}}
\end{figure*}

\vspace{3mm}
{\bf 5)} Amendola-Gannouji-Polarski-Tsujikawa model \\
In the fifth model that we consider, the $f(R)$ is in the form \cite{luca,agp}
\be \label{ag}
f(R)=R-\mu R_*(R/R_*)^p\,.
\ee
This viable model contains three parameters $0<p<1$, $\mu>0$ and $R_*$ which takes a small positive value. The same analysis by setting $p=1/2,\,\mu=5$ and $n=2.75$, shows that $F$ in (\ref{nogo}) becomes negative and the obtained wormhole solution satisfies NEC at the throat (figure \ref{agp}.a). It is also obvious in Fig(\ref{agp}.b) that there are asymptoticly flat traversable wormhole solutions in this model which respect NEC outside the throat $r_0$, if one choose adequate values for $n$.

\begin{figure*}[ht]
\begin{picture}(0,0)(0,0)
\put(120,-217){\footnotesize Fig (7.a)}
\put(355,-217){\footnotesize Fig (7.b)}
\end{picture}
\center
\includegraphics[height=70mm,width=75mm]{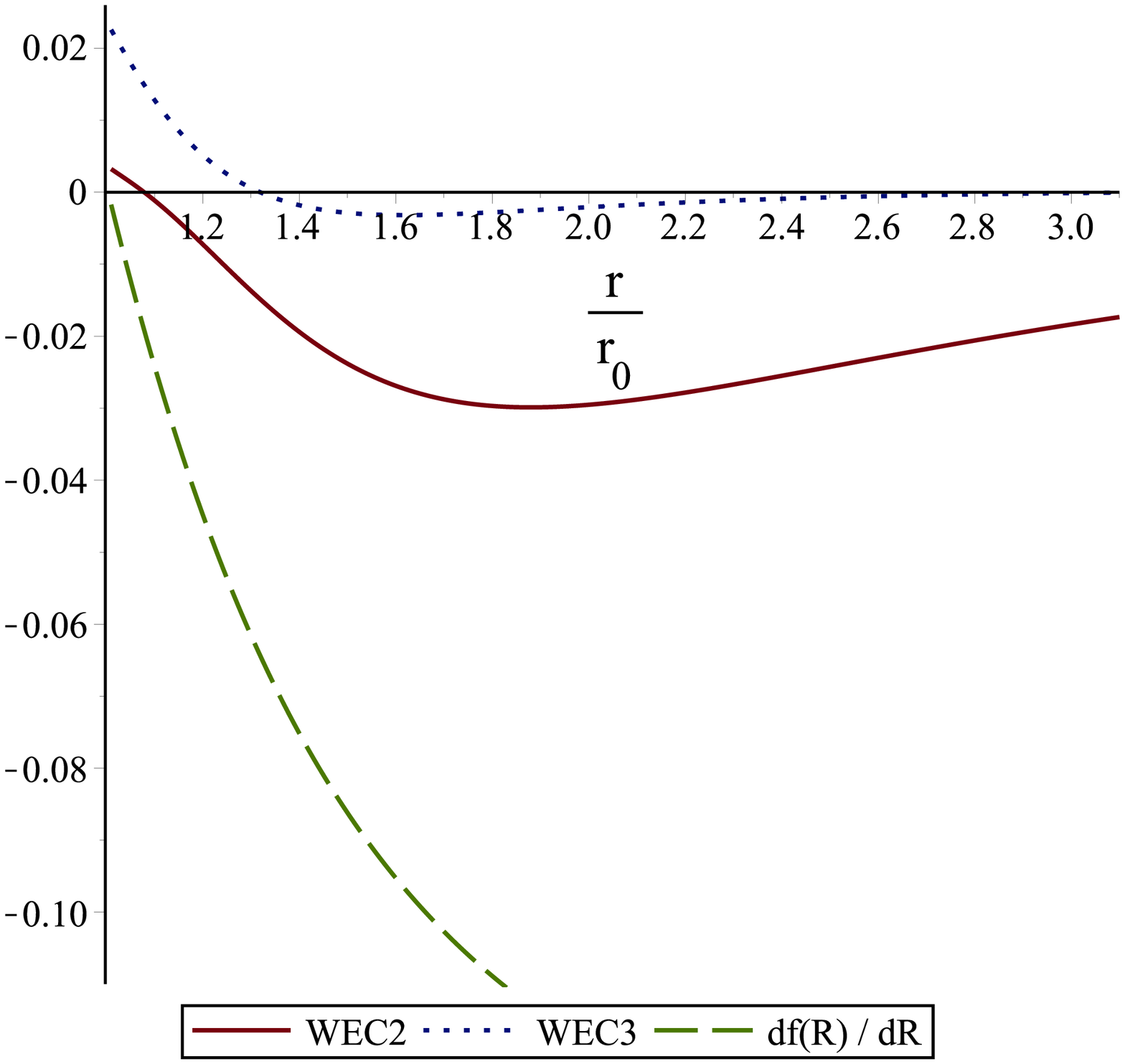} \quad \includegraphics[height=70mm,width=75mm]{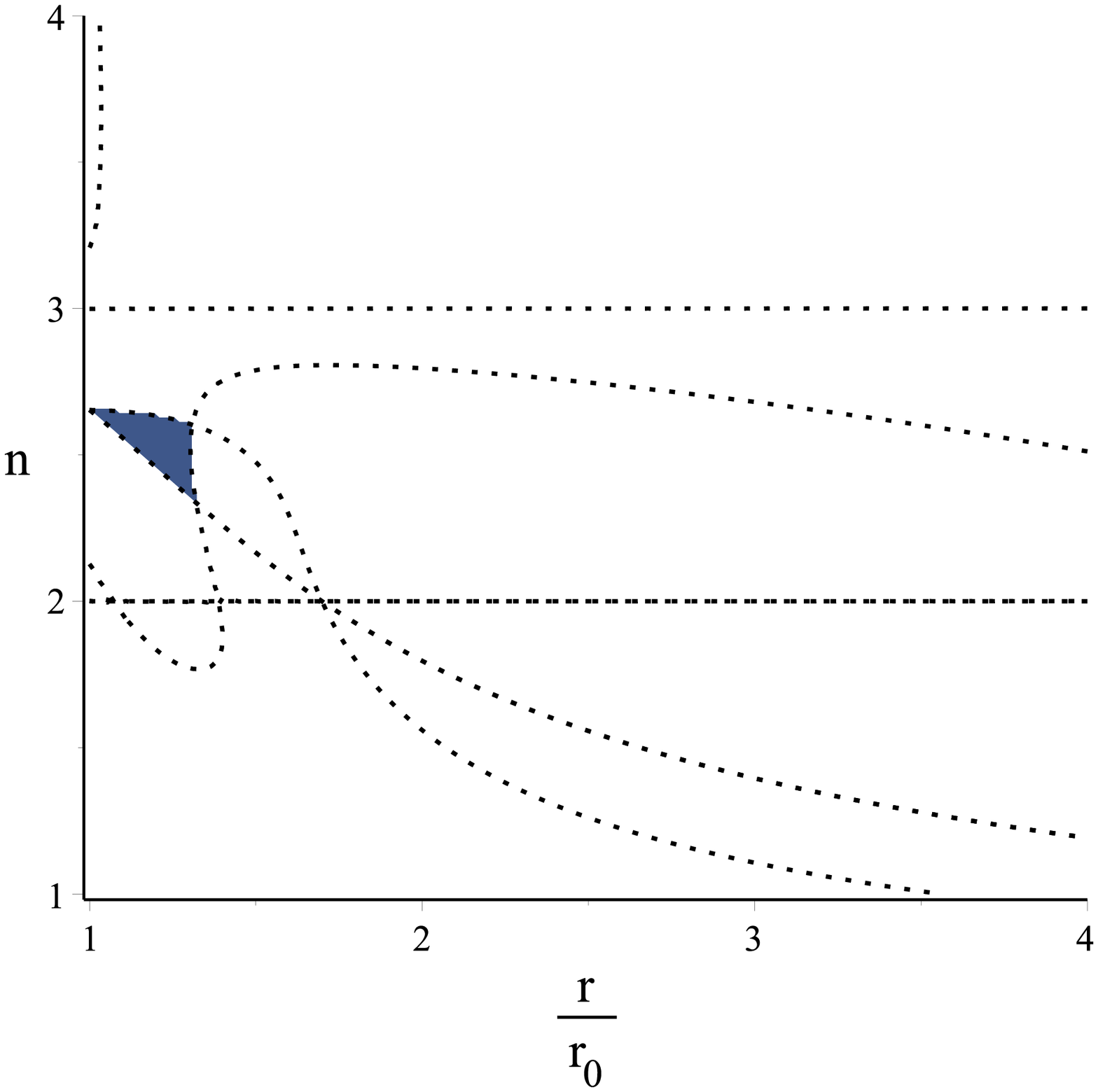}
\vspace{8mm}
\caption{{\small Setting $n=2.65$ in Fig(7.a) we found a wormhole solution in the exponential gravity model that respect NEC. The blue region in Fig(7.b) also shows that it is possible to find wormholes solutions in this model that respect NEC around the throat. In these figures we set $\lambda=2, R_*=0.01$ and $r_0=10$.}\label{exp}}
\end{figure*}

\vspace{3mm}
{\bf 6)} Exponential gravity model \cite{Cognola:2007zu,Elizalde:2010ts}\\
For the last model that we consider in this section, the $f(R)$ is in the form
\be \label{exp1}
f(R)=R-\lambda R_*\left[1-\exp(-\frac{R}{R_*})\right]\,,
\ee
where $\lambda, R_*$ are free positive parameters of the model. Similar to the previous cases, one can find a wormhole in this model which respect NEC at the throat (figure \ref{exp}.a). In figure \ref{exp}.b the $n-r/r_0$ diagram is depicted. It is clear that there is a little possibility for traversable asymptotically flat wormhole solutions in this model to satisfy the null energy conditions around the throat $r_0$.

\subsection{The cases of $c_1\neq 0$}
We considered traversable wormhole solution as (\ref{wmetric}), where $b(r)/r$ given by (\ref{bff}). Remember that the wormhole shape function $b(r)$ satisfies the conditions in (\ref{123}) which leads to $n>2$ in the case of $c_1=-1$ and $n>2.4$ in the case of $c_1=1$, therefore the second term in (\ref{bff}) falls down at large $r$. In this subsection we choose $c_1=\pm 1$,  for which the wormhole metric (\ref{wmetric}) 
at large $r$ matches  hyperbolic ($c_1=-1$) and spherical ($c_1=1$) FRW metric, so we call them asymptotically hyperbolic and asymptotically  spherical wormhole solutions respectively. In the following we verify the null and weak energy condition for these asymptotically  hyperbolic and spherical wormhole solutions in the background of static $f(R)$ modified gravity models that we considered in the previous subsection.

\vspace{3mm}
{\bf 1)} The Tsujikawa model

\begin{figure*}[ht]
\begin{picture}(0,0)(0,0)
\put(139,-192){\footnotesize Fig (8.a)}
\put(338,-192){\footnotesize Fig (8.b)}
\put(139,-363){\footnotesize Fig (8.c)}
\put(338,-363){\footnotesize Fig (8.d)}\end{picture}
\center
\includegraphics[height=61mm,width=65mm]{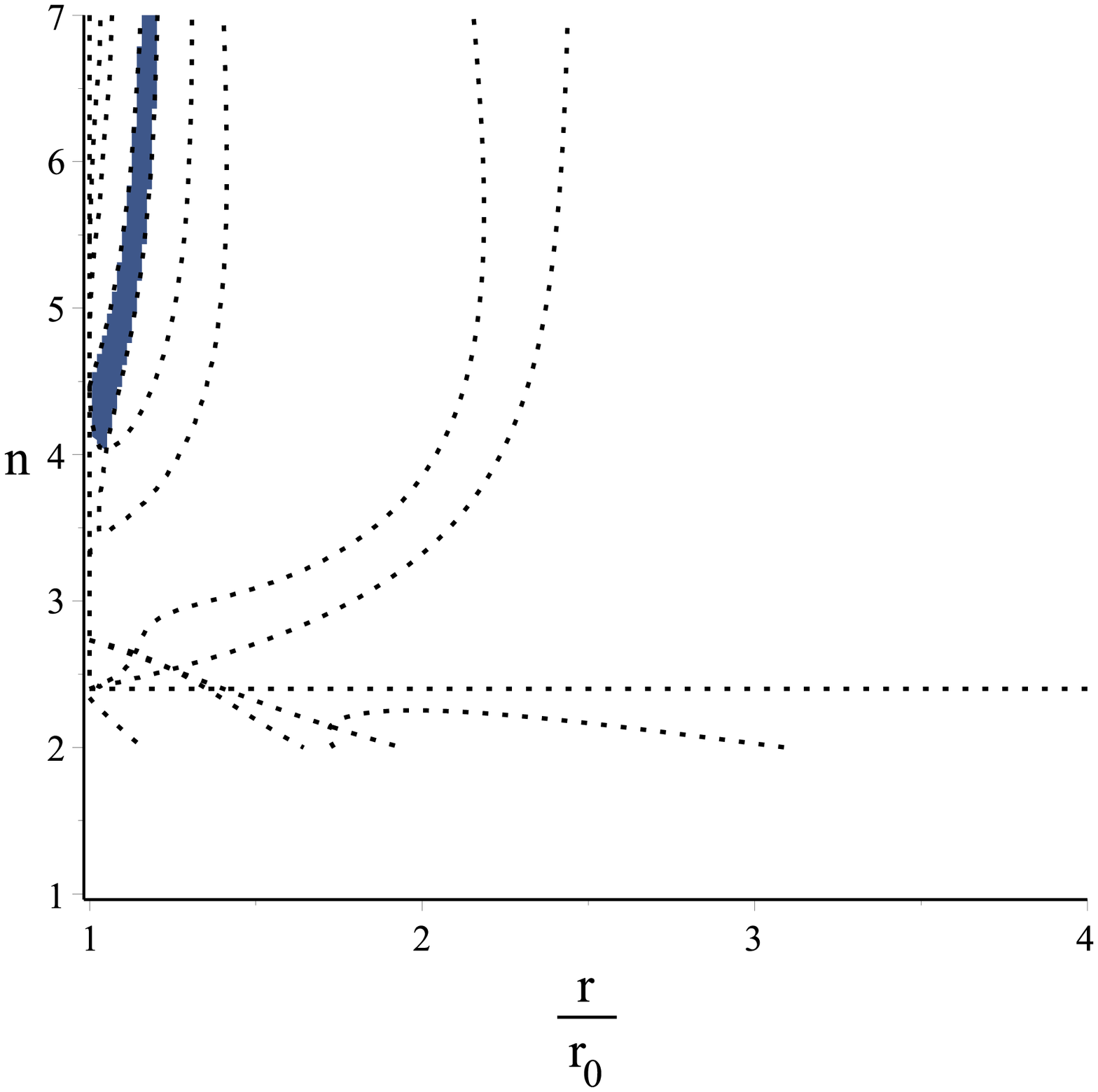}\qquad \includegraphics[height=61mm,width=65mm]{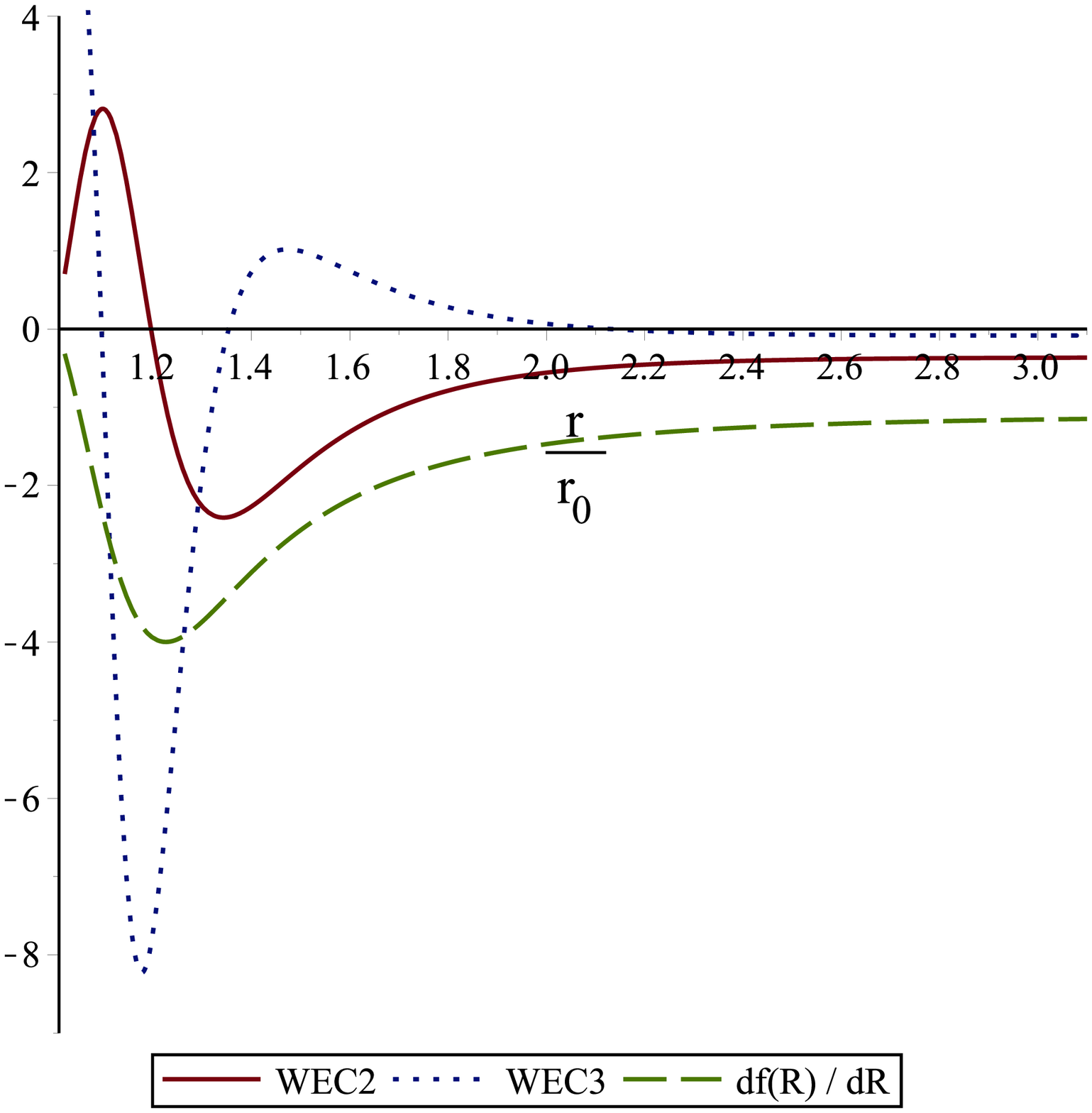}\\
 \vspace{5mm} \includegraphics[height=56mm,width=65mm]{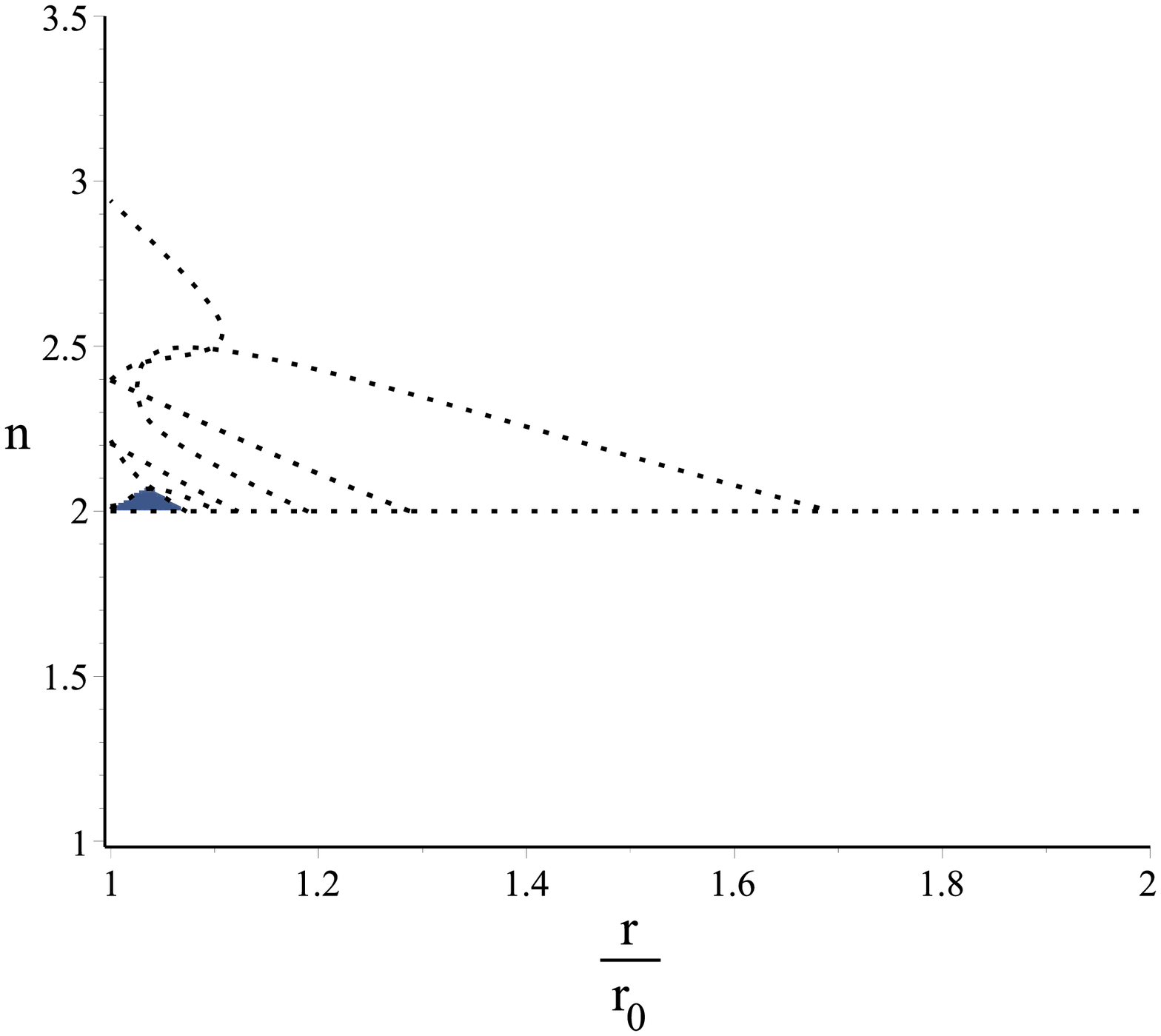}
\qquad \includegraphics[height=56mm,width=65mm]{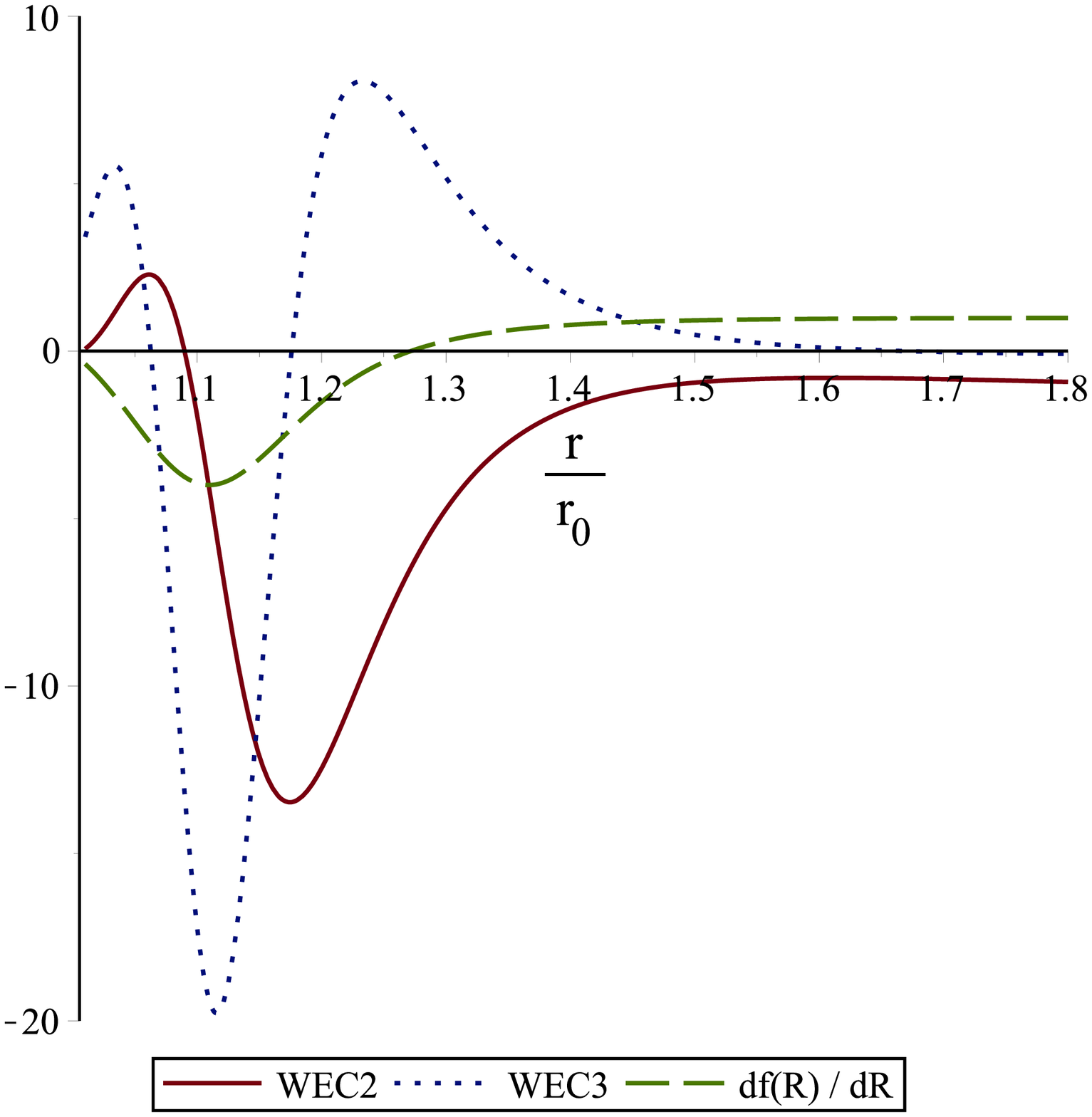}
\vspace{8mm}
\caption{{\small The blue region in Fig8.a (Fig8.c) shows the asymptotically spherical (hyperbolic) wormhole solutions of the Tsujikawa model that respect the NEC. Fig(8.b) which is depicted by setting $n=4.5, r_0=1$, shows that $F=df(R)/dR<0$ and the asymptotically spherical wormhole in this model respect NEC around the throat. Fig(8.d) demonstrates that for an asymptotically open wormhole with $n=2.02, r_0=0.5$, NEC is respected around the throat. In these figures we set $\mu=5, R_*=1$.}\label{tsu1}}
\end{figure*}

 For this model the $f(R)$ is given in (\ref{tsu}). We are interested in solutions for which $F$ in (\ref{nogo}) takes negative values, conditions {\it ii}), {\it iii} in (\ref{123}) are satisfied and (\ref{wec}) are respected simultaneously. Figures \ref{tsu1}.a,c show the location of these asymptotically closed ($c_1=1$) and open ($c_1=-1$) wormhole solutions in the $n-r/r_0$ diagram respectively.  In figure \ref{tsu1}.b (\ref{tsu1}.d) we have explicitly  depicted $F$ and requirements of the WEC for an asymptotically spherical ( hyperbolic) wormhole solution with $n=4.5$ ($n=2.02$). It is obvious that by choosing adequate values for the parameters, one can find traversable asymptotically closed and open wormhole solutions in this $f(R)$ gravity model that respect NEC around the wormhole throat.

\vspace{4mm}
{\bf 2)} Hu-Sawicki model

\begin{figure*}[ht]
\begin{picture}(0,0)(0,0)
\put(142,-192){\footnotesize Fig (9.a)}
\put(339,-192){\footnotesize Fig (9.b)}
\put(142,-363){\footnotesize Fig (9.c)}
\put(339,-363){\footnotesize Fig (9.d)}\end{picture}
\center
\includegraphics[height=61mm,width=65mm]{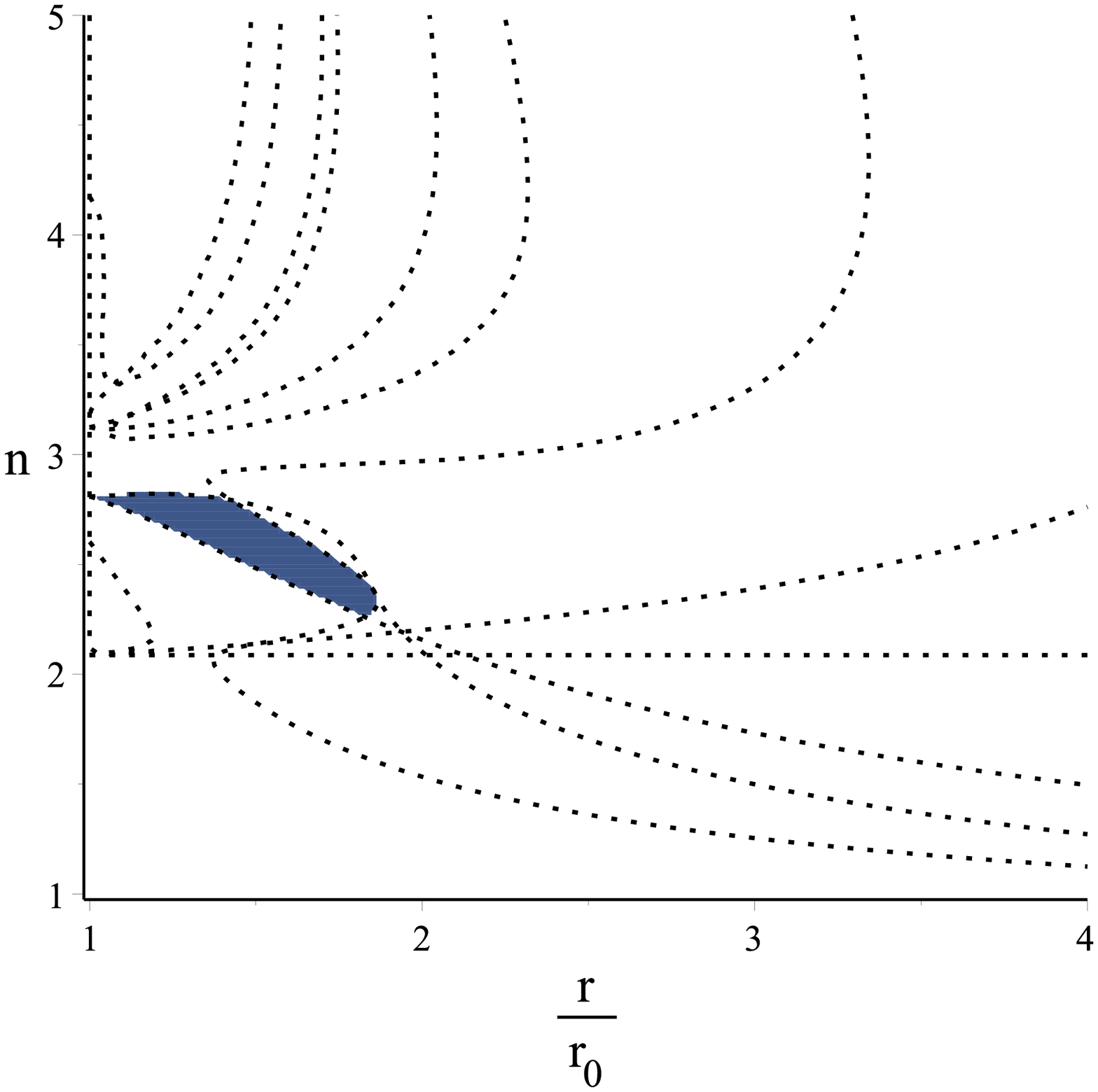}\qquad \includegraphics[height=61mm,width=65mm]{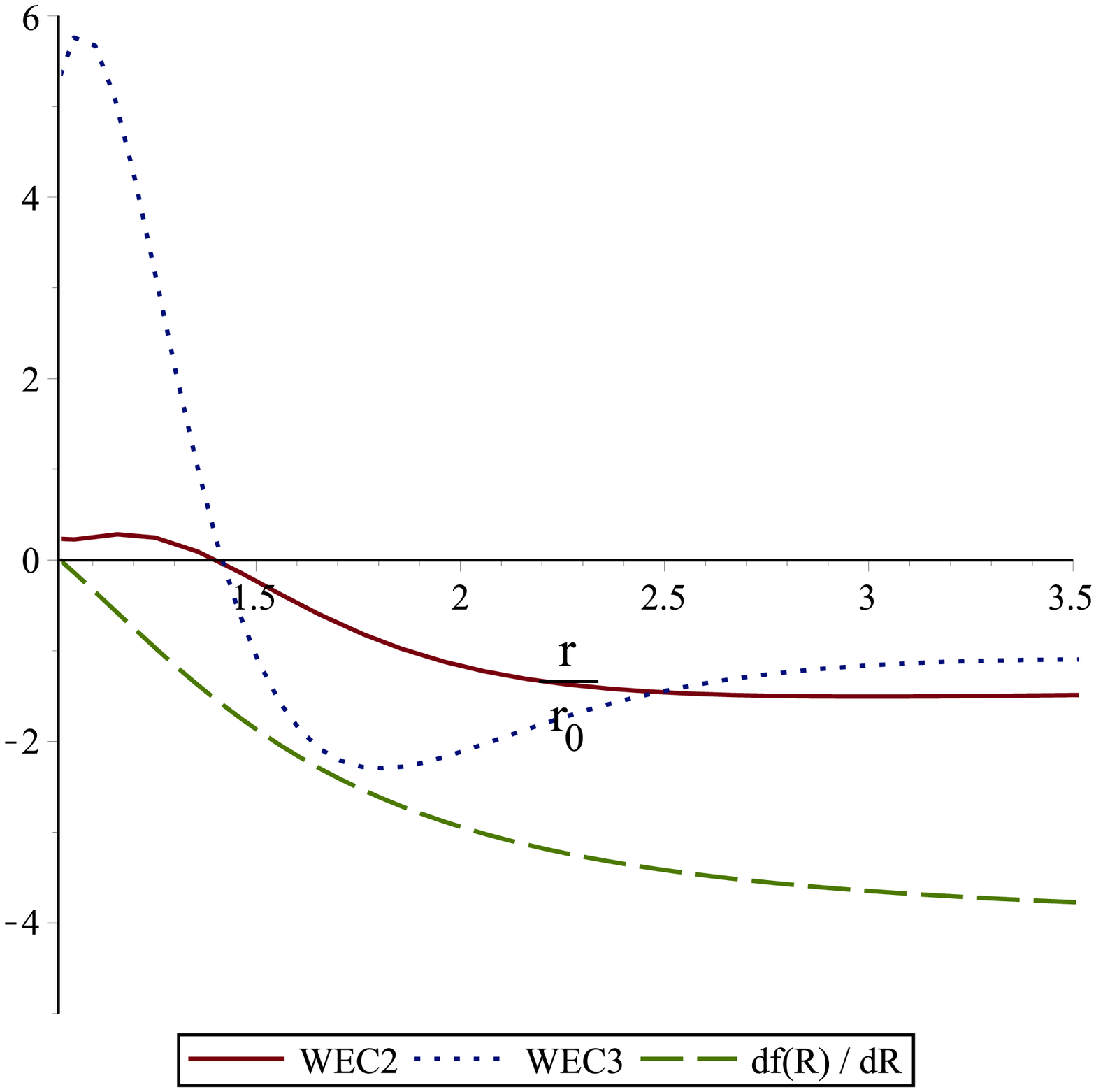}\\
 \vspace{5mm} \includegraphics[height=56mm,width=65mm]{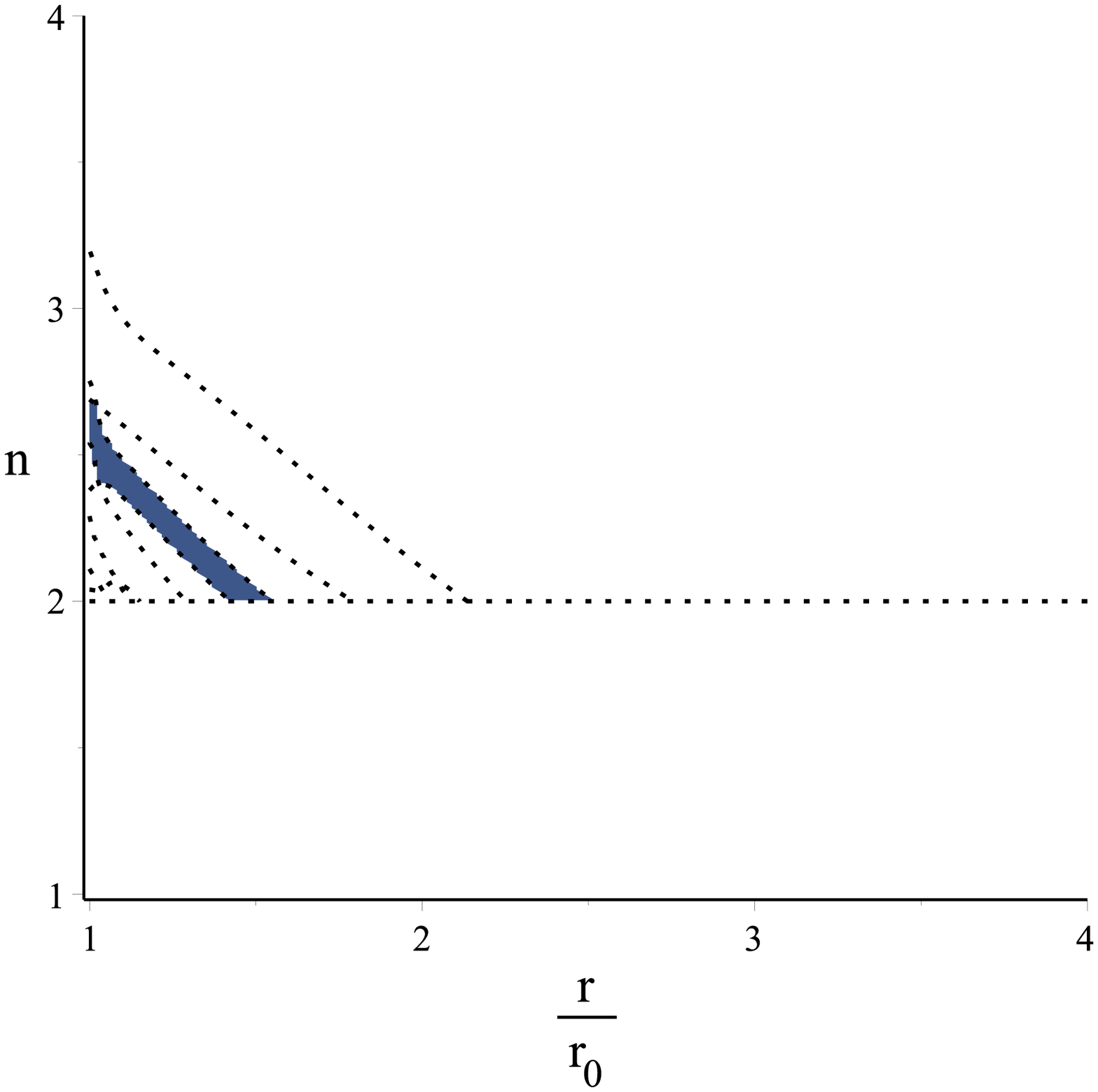}
\qquad \includegraphics[height=56mm,width=65mm]{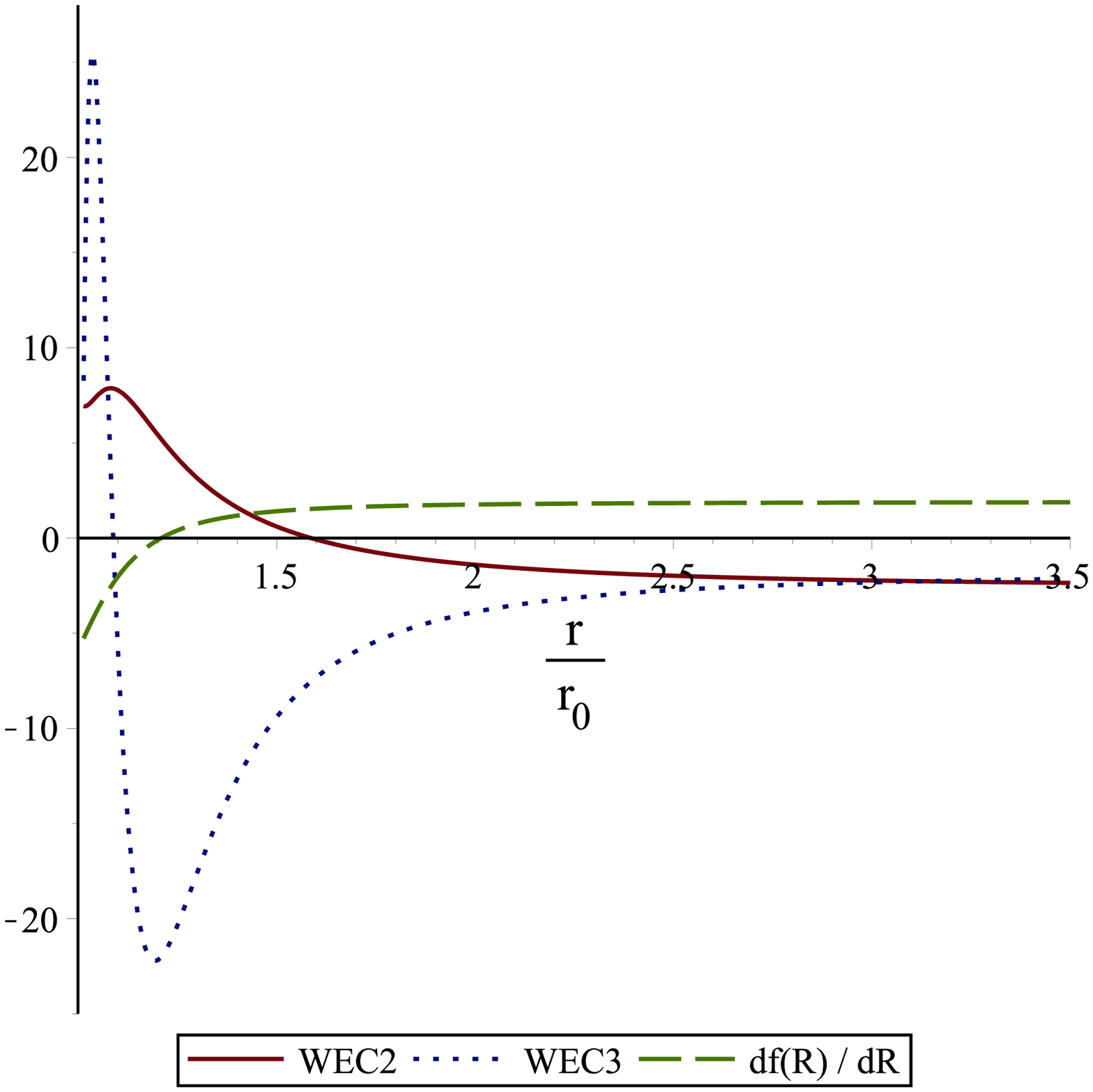}\vspace{8mm}
\caption{{\small Similar to the previous model, there are asymptotically spherical (Fig(9.a)) and hyperbolic (Fig(9.c)) wormholes in Hu-Sawicki model that respect the NEC. In Fig(9.b) we draw $F$ and WEC2,3 for an asymptoticly spherical solution with $n=2.8$. Fig(9.d) also shows that an asymptoticly  hyperbolic wormhole with $n=2.5$, respect the NEC. In these figures we set the parameters as $\mu=10, R_*=1, r_0=.5$.}\label{hs2}}
\end{figure*}

The second model that we studied its wormhole solutions at the previous subsection was the Hu-Sawicki model with the $f(R)$ in the form (\ref{hs}). We verify the NEC for  asymptotically  spherical and hyperbolic wormhole solutions of this model in Fig(\ref{hs2}a,c) respectively. In the blue regions in these figures $F=df(R)/dR$ becomes negative, {\it ii}), {\it iii}) of (\ref{123}) are satisfied and (\ref{wec}) are respected simultaneously.
 We have plotted the NEC for an asymptotically spherical and  hyperbolic wormhole solution in figure \ref{hs2}.b,d respectively.
Similar to the previous model, it is clear that by setting the parameters in adequate values, traversable asymptotically  spherical and hyperbolic wormholes in this model respect the null energy condition around the throat $r_0$.

\vspace{3mm}
{\bf 3)} The Starobinsky model

\begin{figure*}[ht]
\begin{picture}(0,0)(0,0)
\put(135,-203){\footnotesize Fig (10.a)}
\put(339,-203){\footnotesize Fig (10.b)}
\end{picture}
\center
\includegraphics[height=65mm,width=65mm]{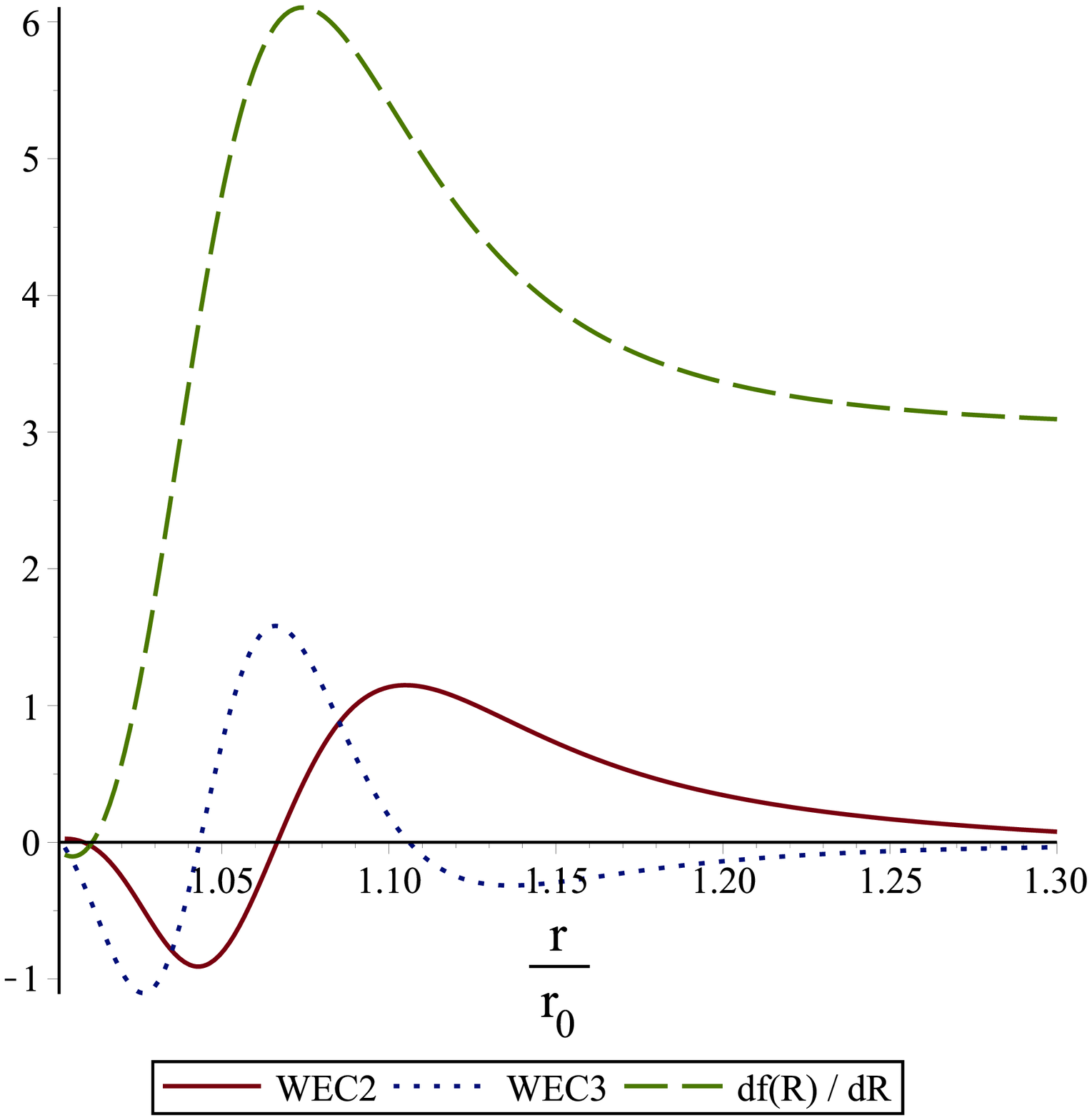}\qquad \includegraphics[height=65mm,width=65mm]{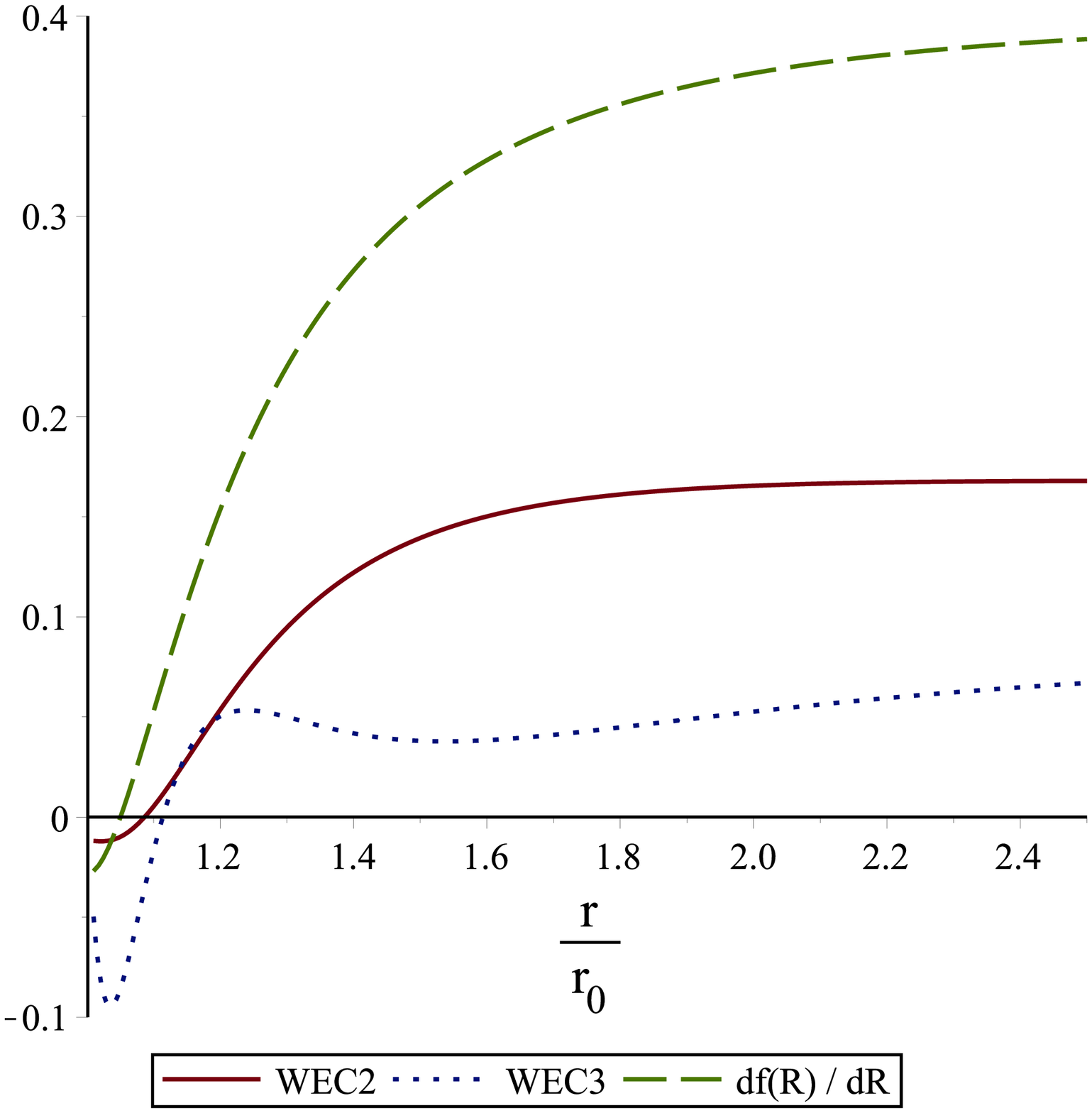}
\vspace{8mm}
\caption{{\small In the case of Starobinsky model, Fig(10.a) and Fig(10.b) show $F$, WEC2,3 for the asymptotically  hyperbolic and spherical wormhole solutions respectively. In both cases by fixing $\lambda=1, R_*=1$ the $F<0$ is satisfied around $r_0$, however the NEC are violated. We set $r_0=0.55, n=2.25$ in (10.a) and $r_0=1.5, n=4$ in (10.b).}\label{star2}}
\end{figure*}

In the case of Starobinsky model which $f(R)$ is given in (\ref{staro}),  the $df(R)/dR<0$ could be satisfied for both asymptotically  spherical and hyperbolic solutions by choosing $R_*$ and $r_0$ appropriately, however the null energy condition is violated around the throat by these wormhole solutions (Fig\ref{star2}). In this model the $n-r/r_0$ diagram is blank which means that there is no wormhole solution in this case which respect the NEC. 

 \vspace{3mm}
 {\bf 4)} Nojiri-Odintsov model
 
\begin{figure*}[ht]
\begin{picture}(0,0)(0,0)
\put(78,-192){\footnotesize Fig (11.a)}
\put(233,-192){\footnotesize Fig (11.b)}
\put(397,-192){\footnotesize Fig (11.c)}
\end{picture}
\center
\includegraphics[height=60mm,width=57mm]{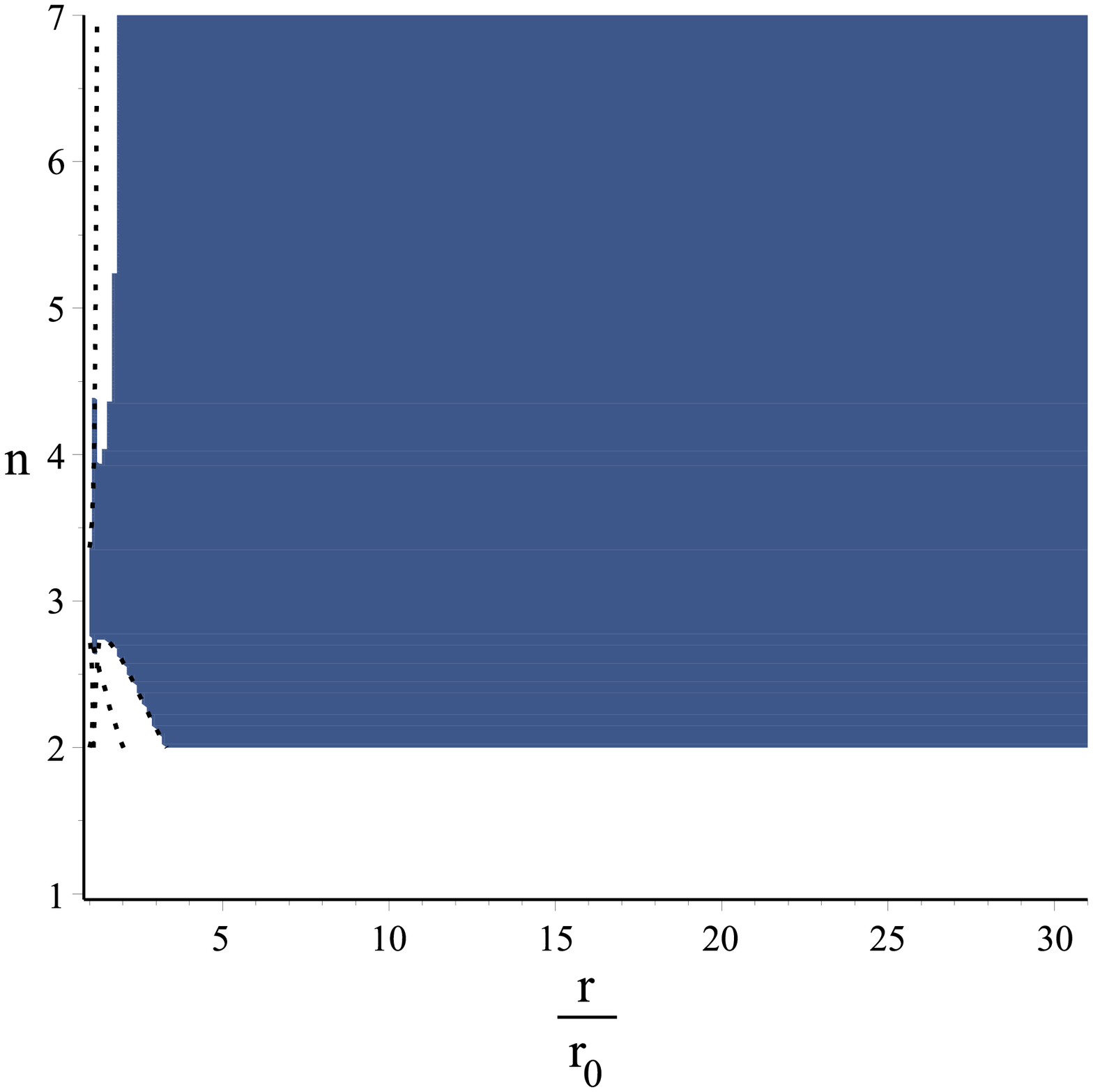}\, \includegraphics[height=60mm,width=56mm]{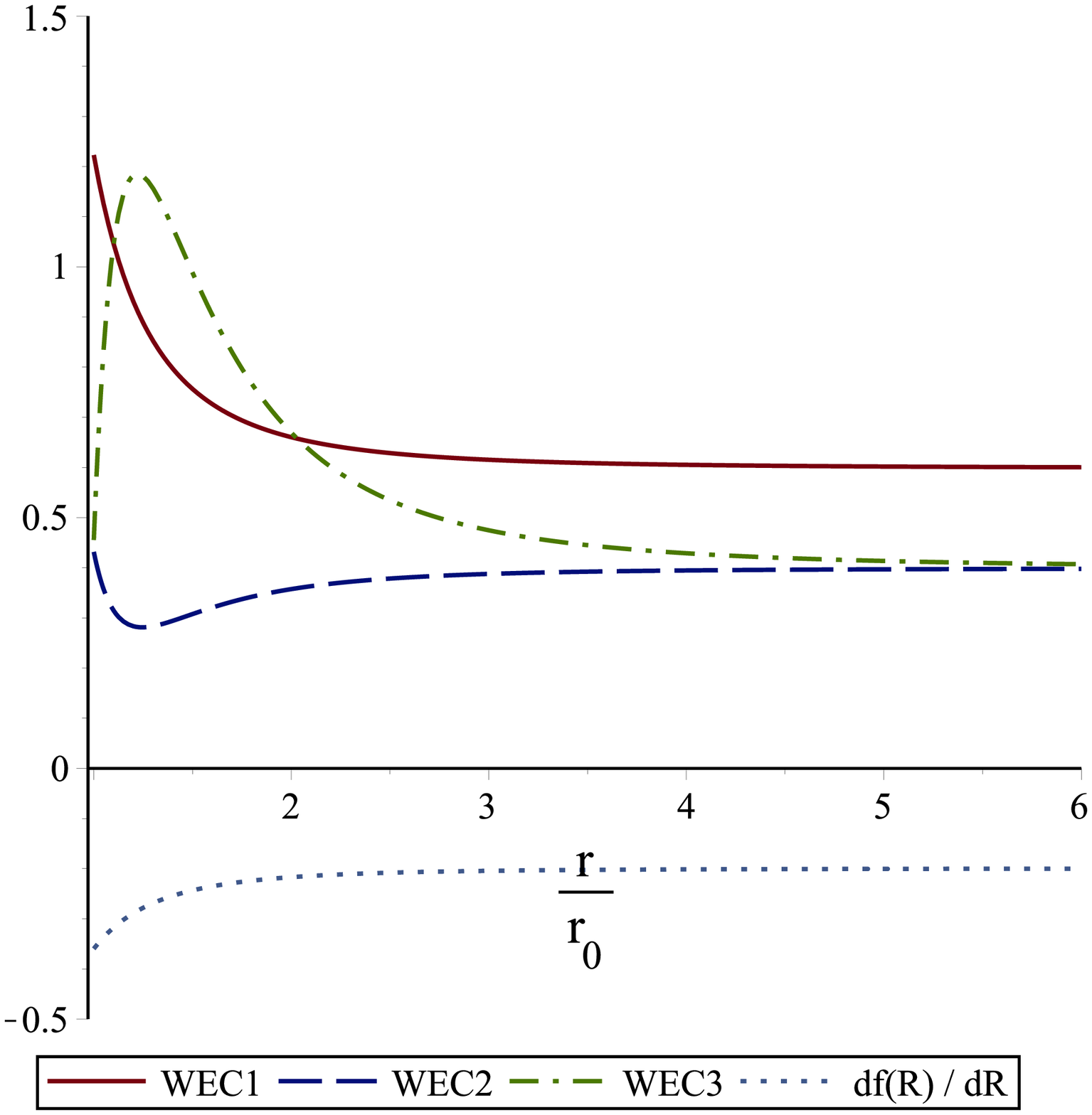}\,
\includegraphics[height=60mm,width=57mm]{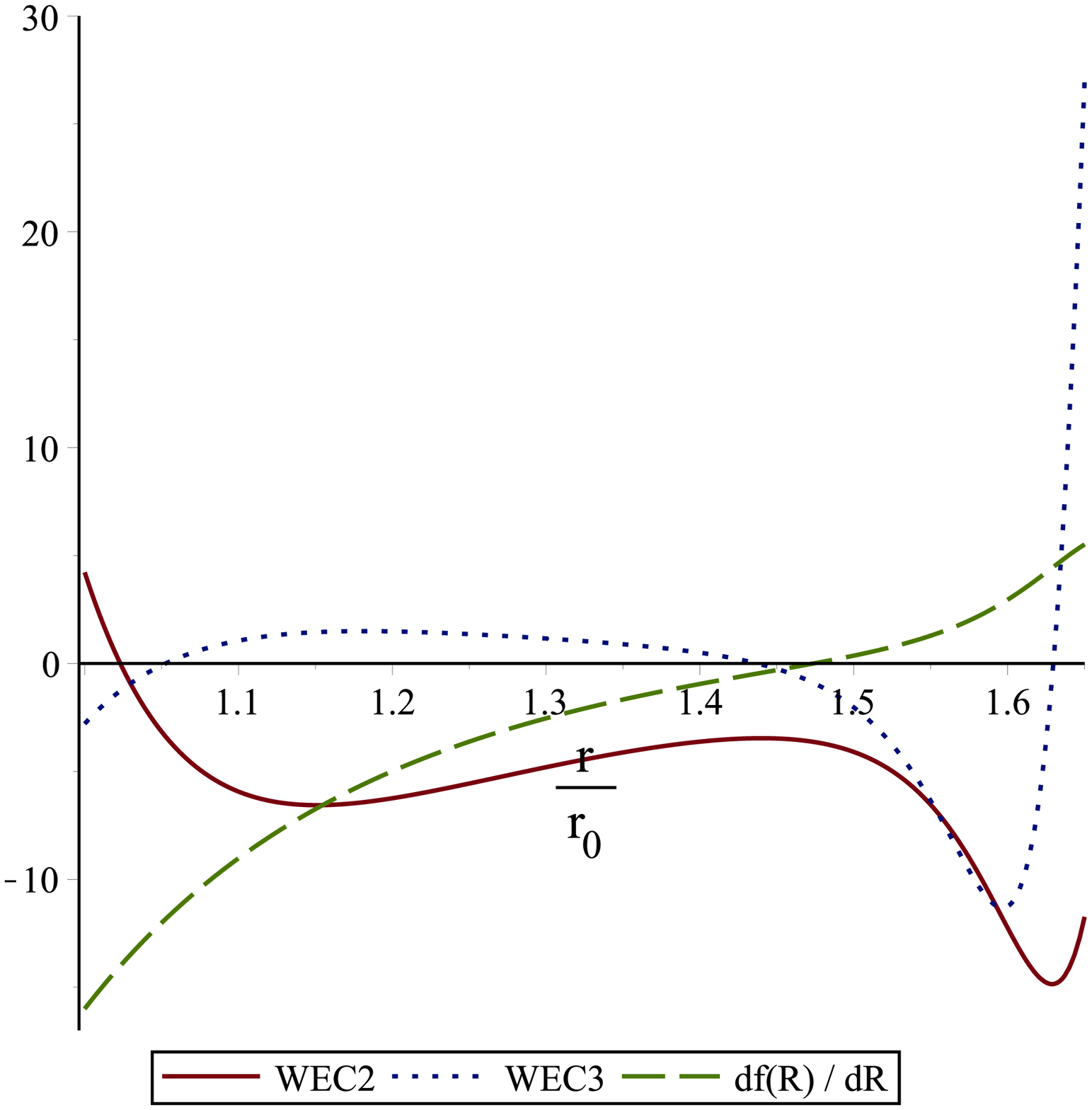}\vspace{8mm}
\caption{{\small Fig(11.a) shows that the asymptoticly  hyperbolic wormholes with $n>2$ in the Nojiri-Odintsov model, respect the WEC almost through the whole space outside the throat $r_0$. We have depicted WEC1,2,3 and $F$ for an asymptotically hyperbolic wormhole with $n=3.2$ in Fig(11.b). The asymptoticly spherical wormholes do not respect the NEC in this background (Fig(11.c)). In Fig(11a,b) we set $a=2,\, b=20,\, c=20$ and $r_0=1$. Figure (11.c) is drawn by setting $a=5,\, b=20,\, c=10$ and $r_0=0.5$\,.}\label{no2}}
\end{figure*}

 Now we return to the Nojiri-Odintsov modified gravity model where the form of $f(R)$ is given in (\ref{no}).  The $n-r/r_0$ diagram for asymptotically hyperbolic ($c_1=-1$) solutions in this model by setting $m=2$, is depicted in Fig(\ref{no2}.a). The blue zone in this diagram corresponds to the region that $F<0$, {\it ii}), {\it iii}) of (\ref{123}) and (\ref{wec}) are satisfied simultaneously. It is clear that the asymptoticly hyperbolic wormholes with $n>2$ in this model, respect the WEC almost through  the whole space outside the throat $r_0$. In figure (\ref{no2}.b) we have drawn $df(R)/dR$ and WEC1,2,3 explicitly for a wormhole solution with $n=3.2$\,.
 
In the case of $c_1=1$, it is possible to set the parameters in a manner that $F<0$ is satisfied, however these asymptotically  spherical wormhole solutions do not respect the NEC (see Fig(\ref{no2}.c) for instance), in the other words, the $n-r/r_0$ diagram in this case is blank.

\vspace{3mm}
{\bf 5)} Amendola-Gannouji-Polarski-Tsujikawa model

\begin{figure*}[ht]
\begin{picture}(0,0)(0,0)
\put(79,-203){\footnotesize Fig (12.a)}
\put(231,-203){\footnotesize Fig (12.b)}
\put(397,-203){\footnotesize Fig (12.c)}
\end{picture}
\center
\includegraphics[height=65mm,width=57mm]{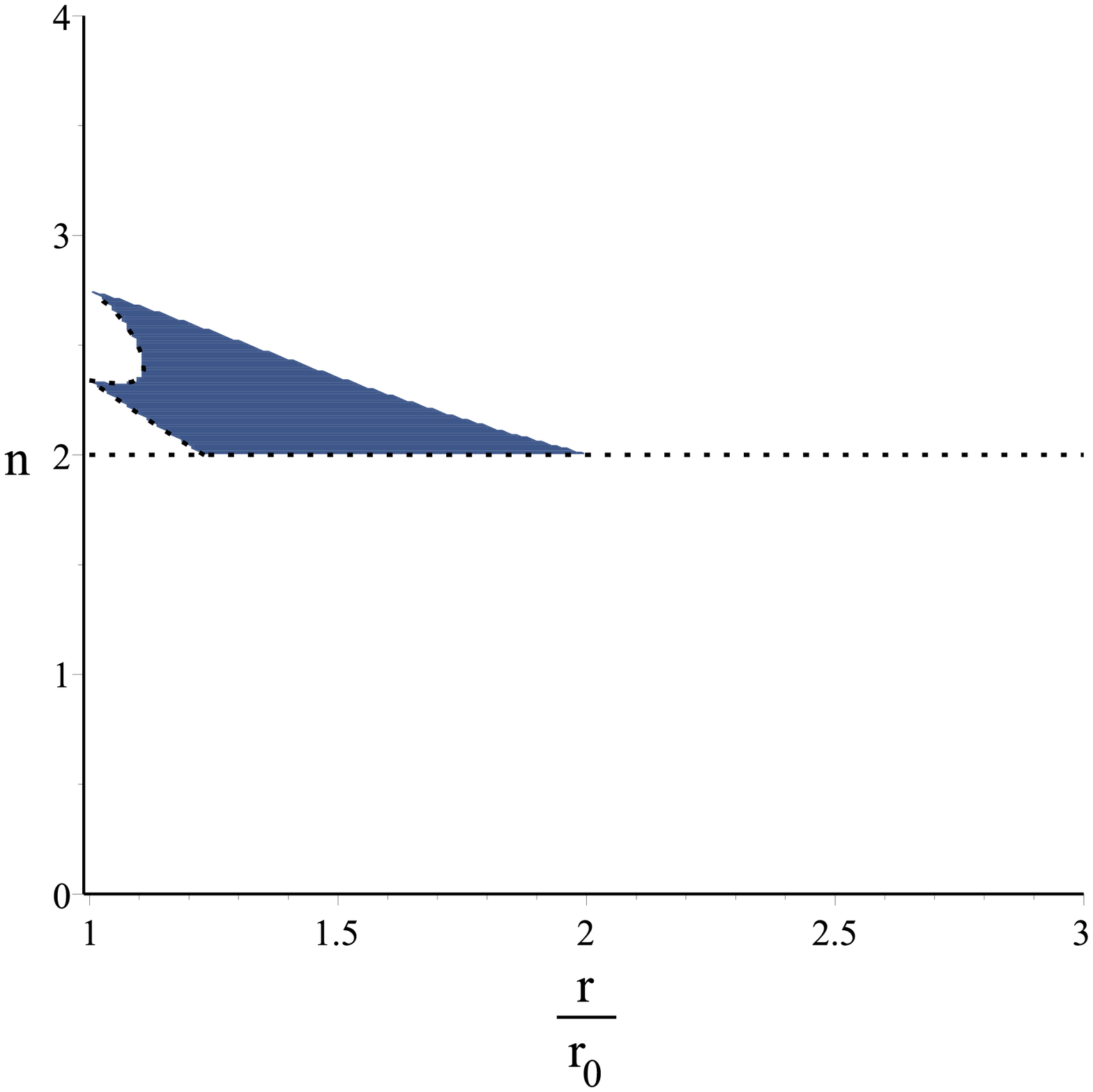}\,\includegraphics[height=65mm,width=57mm]{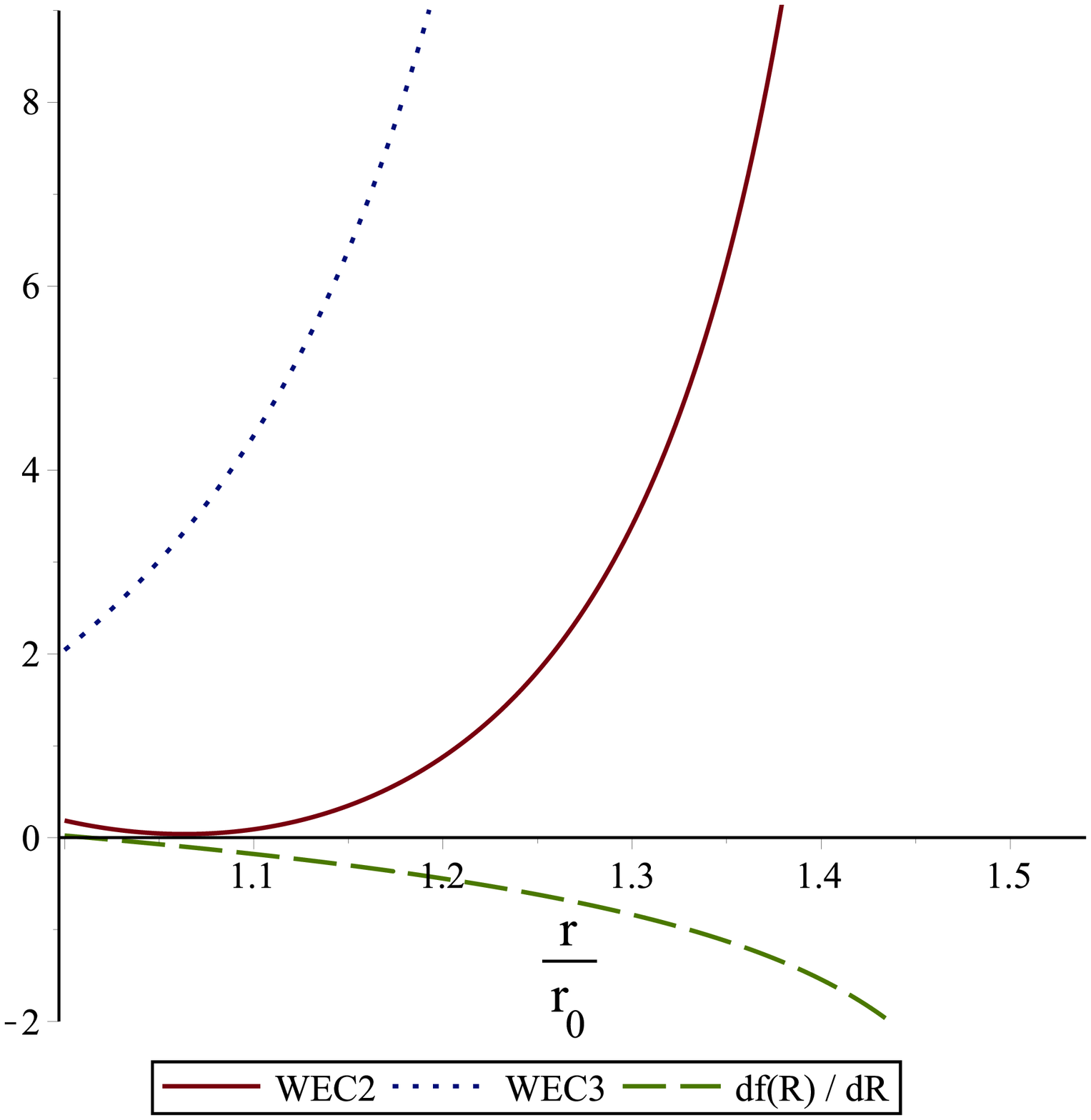}\,\includegraphics[height=65mm,width=58mm]{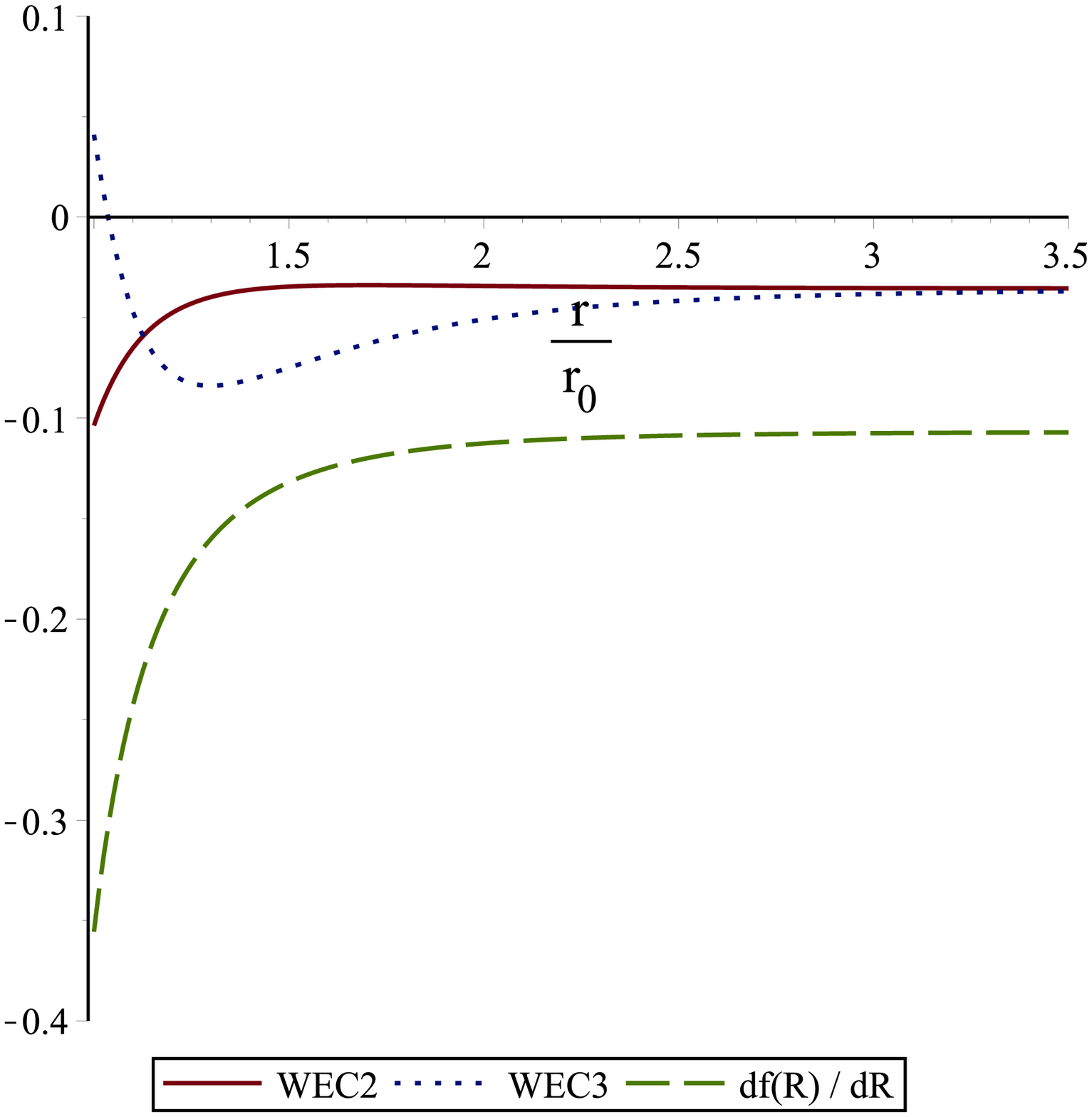}
\vspace{6mm}
\caption{{\small The blue region in Fig(12.a) corresponds to asymptoticly hyperbolic wormhole solutions that respect the NEC in Amendola-Gannouji-Polarski-Tsujikawa model. It is clear from Fig(12.b) that $F<0$ and NEC is respected around $r_0$, for an asymptoticly hyperbolic wormhole with $n=2.3$\,. In Fig(12.c) by setting $n=5$ we show an example for violation of the NEC in the case of asymptotically spherical wormholes in this model. We set the parameters as $p=1/2\,, R_*=0.1, \mu=20, r_0=0.3\, (\mu=7, r_0=2)$ in Fig12a,b (Fig12.c).}\label{agp2}}
\end{figure*}

The $f(R)$ in this model is given by (\ref{ag}). In the case of $c_1=-1$ we check the NEC for wormhole solutions of this model in figure (\ref{agp2}.a) which shows that it is possible to find asymptotically hyperbolic wormholes that respect NEC around the throat, by choosing appropriate parameters in this static model. In figure (\ref{agp2}.b) we have explicitly drawn the NEC for a wormhole by setting $n=2.3$\,.

In the case of $c_1=1$ our results show that, like the previous model, there is no asymptoticly closed wormhole solution in this model which respect NEC (the $n-r/r_0$ diagram is blank). As an example we have depicted $F=df(R)/dR$ and WEC2,3 for a wormhole by setting $n=5$ in figure (\ref{agp2}.c).

\vspace{3mm}
{\bf 6)} Exponential gravity model

\begin{figure*}[ht]
\begin{picture}(0,0)(0,0)
\put(78,-192){\footnotesize Fig (13.a)}
\put(230,-192){\footnotesize Fig (13.b)}
\put(402,-192){\footnotesize Fig (13.c)} 
\end{picture}
\center
\includegraphics[height=60mm,width=59mm]{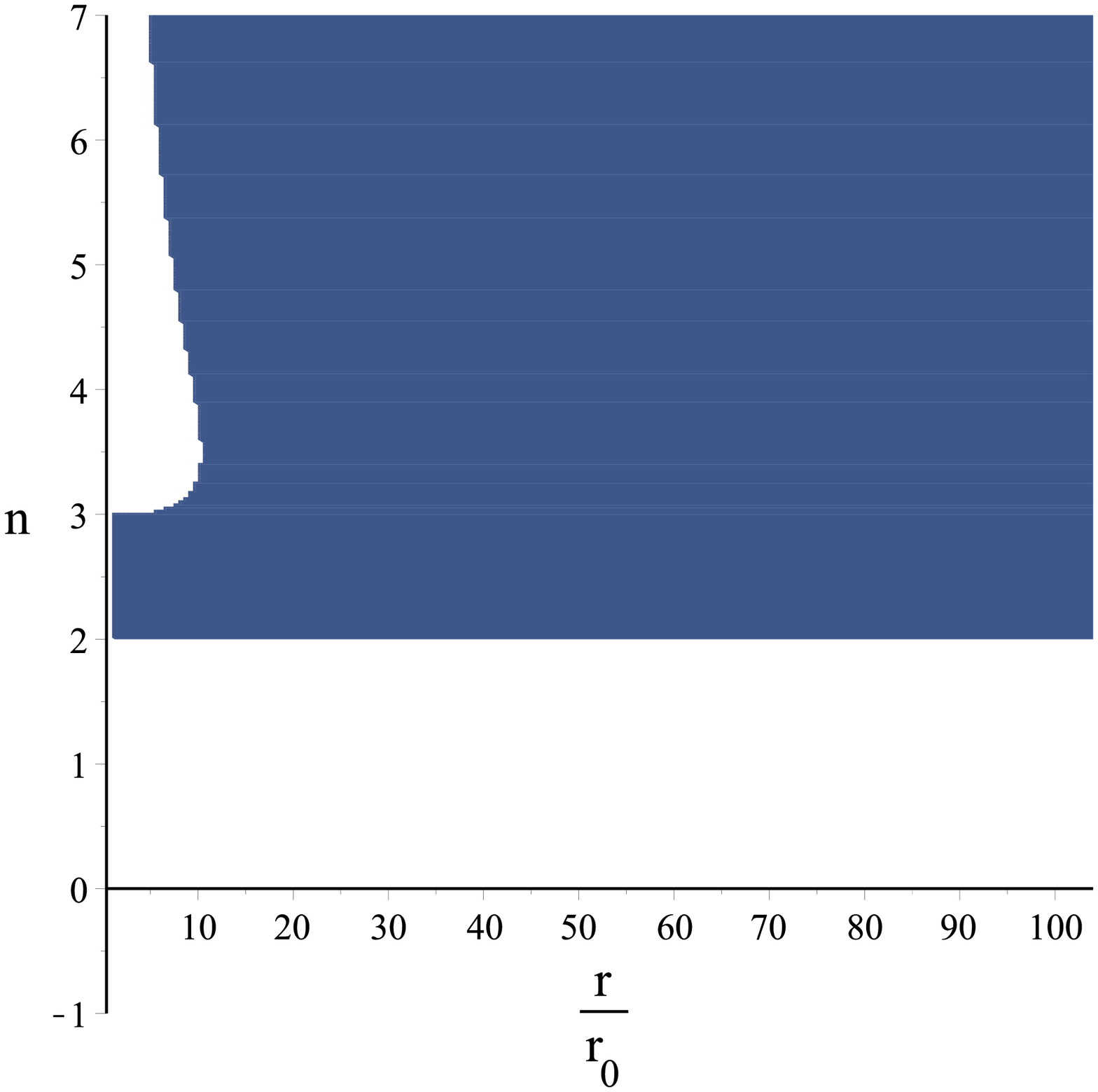}\, \includegraphics[height=60mm,width=57mm]{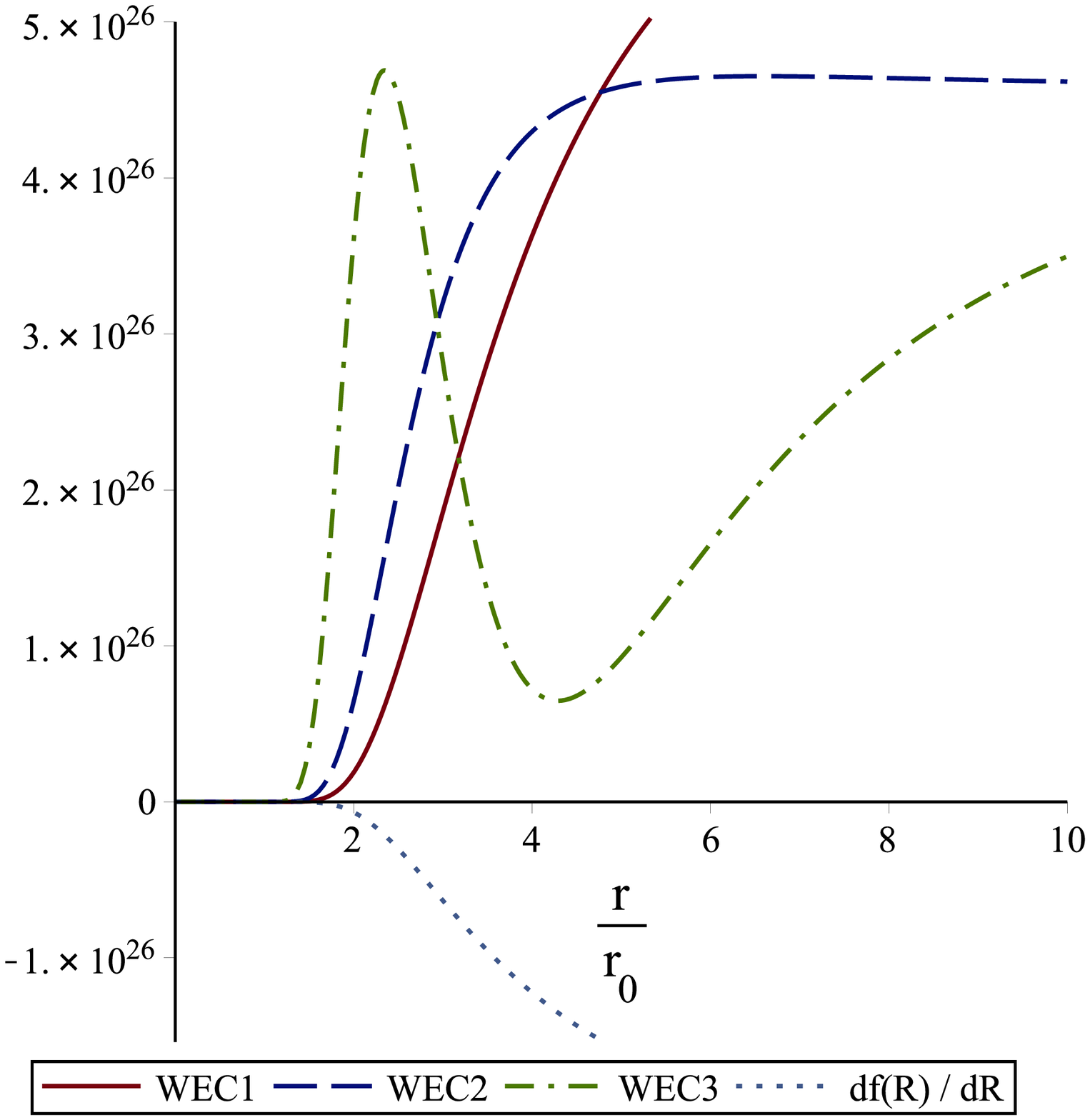} \, \includegraphics[height=60mm,width=58mm]{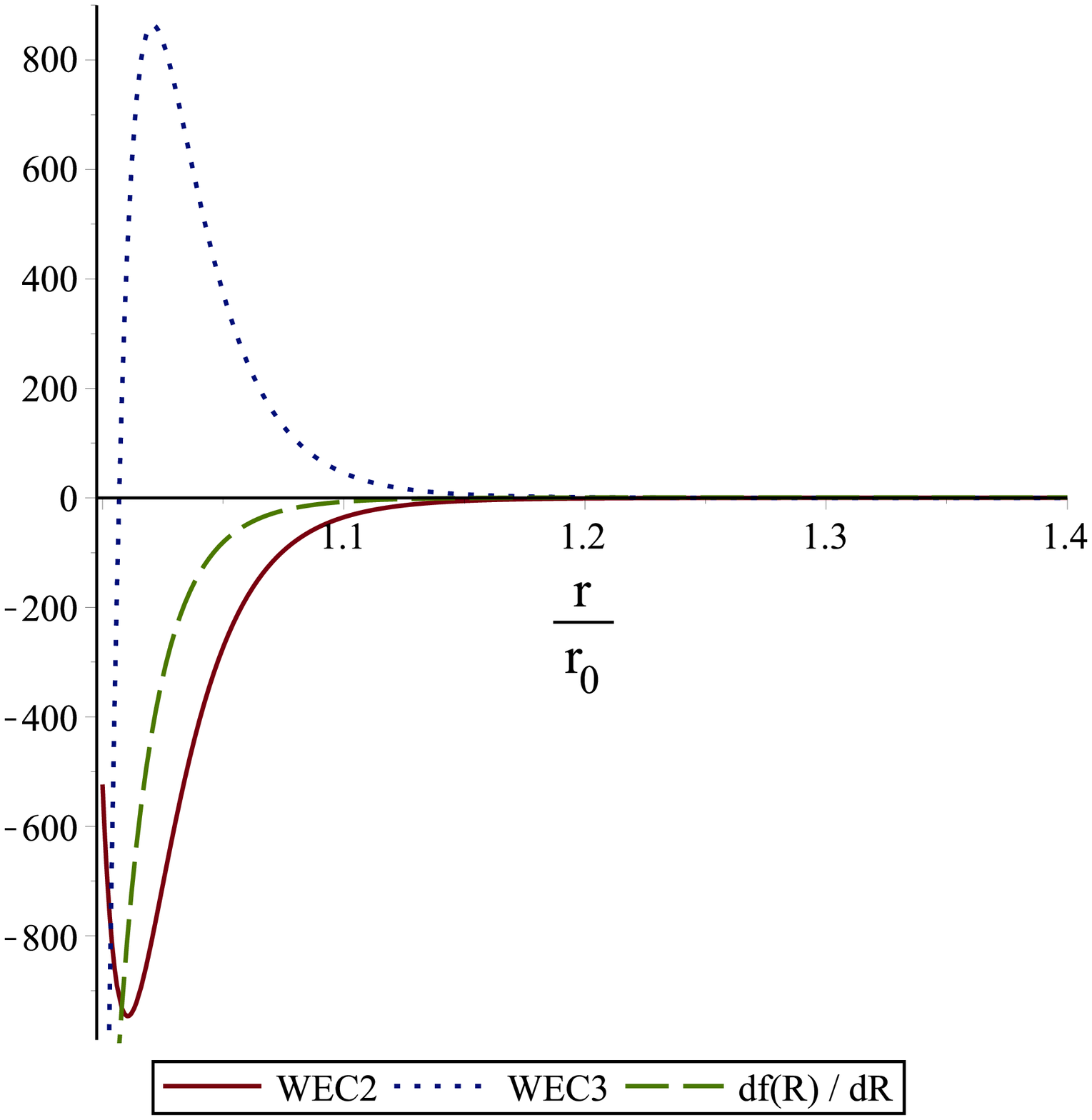}
\vspace{6mm}
\caption{{\small In the case of $c_1=-1$, Fig(13.a) shows the region that $F<0$ and (\ref{wec}) are satisfied in blue. This means that the WEC is respected by the asymptotically hyperbolic wormhole solutions of the exponential gravity model, almost through the whole space outside the throat. In Fig(13.b) we have plotted $F$ and WEC1,2,3 for an asymptoticly hyperbolic wormhole with $n=2.5$. In the case of $c_1=1$, Fig(13.c) shows that the NEC is  violated at the throat for a wormhole with $n=4$. In these figures we set $\lambda=2,\, R_*=0.1$ and $r_0=1$.}\label{exp3}}
\end{figure*}

For the last model that we considered in the previous subsection, the $f(R)$ is given by (\ref{exp1}).  The same analysis shows that in the case of $c_1=-1$, there are asymptotically  hyperbolic wormhole solutions which respect the WEC almost through the whole space outside the wormhole throat (figure \ref{exp3}.a,b). It is also possible in the case of $c_1=1$, to find asymptoticly  spherical wormhole solutions by choosing parameters in a manner that $F<0$. However the NEC are violated by these wormholes as we can see in the Fig(\ref{exp3}.c).

\section{Conclusion}
In this paper we considered wormhole solutions in the context of some $f(R)$ modified gravity models. In the first step, by solving the field equations similar to \cite{Lobo:2009ip}, we found the energy density, radial and transverse pressure (\ref{ro})-(\ref{pt}) of the solutions in terms of $F=df(R)/dR$ and the shape function $b(r)$. In the second step we considered the Ricci scalar of the wormhole solutions in the form (\ref{expand}) and reorganized the shape function into the form $b(r)=\left[\left( -r_0^{n}\,c_1+r_0^{n-2} \right) {r}^{3-n}+{c_1\,r^3} \right]$.

As the next step we inserted the physical conditions on $b(r)$ and found that $n$ must be greater than 2. This convey us to choose the values $c_1=-1,0,+1$ for which the wormhole solution (\ref{wmetric}) at large radial coordinate $r$, matches the hyperbolic, flat and spherical FRW universe respectively (asymptotically  hyperbolic, flat and spherical solutions).
In the final step we chose six  static $f(R)$ modified gravity models and checked whether the null and weak energy conditions (\ref{wec}) hold for the wormhole solutions in the background of these $f(R)$ models.

 Our results show that by choosing adequate parameters, it is possible to find traversable asymptotically hyperbolic and flat wormhole solutions that respect NEC around the throat $r_0$, in all these $f(R)$ models except that the Starobinsky model. In the case of the Tsujikawa model, asymptoticly flat wormholes can also respect the weak energy condition through the whole space outside the throat. In the case of exponential gravity and Nojiri-Odintsov models, asymptoticly hyperbolic wormholes can respect WEC in the whole space outside the $r_0$ too. We also found that asymptotically spherical wormhole solutions respect the NEC around $r_0$ just in the Tsujikawa and Hu-Sawicki models. It is worth to mention that due to $F=df(R)/dR<0$ for these wormhole solutions, the ``remote mouths'' would be located in anti-gravitational regions because of the effective gravitational constant is negative \cite{Bronnikov:2006pt,Bronnikov:2010tt}.


As a final remark we emphasis that unlike the Einstein gravity, our results show that  it is possible to find traversable wormhole solutions which respect the null and weak energy conditions in the background of some $f(R)$ modified gravity models. This means that $f(R)$ gravity in addition to its applications in cosmology, can be a useful framework in wormhole physics.  


\section*{Acknowledgment}
H.G. would like to thank M.M. Sheikh-Jabbari, Ghadir Jafari and Davood Mahdavian Yekta for useful discussions.




\begin{thebibliography}{99}
\bibitem{fl} L. Flamm, Phys. Z. 17, 448 (1916).
 	\bibitem{eros} A. Einstein and N. Rosen, Phys. Rev. 48, 73 (1935).
 	\bibitem{mtor} M. S. Morris and K. S. Thorne, Am. J. Phys. 56, 395 (1988);
 	M. S. Morris, K. S. Thorne, and U. Yurtsever, Phys. Rev.
 	Lett. 61, 1446 (1988).
 	\bibitem{kar2}S. Kar and D. Sahdev, Phys. Rev. D 53, 722 (1996); A. V. B.
 	Arellano and F. S. N. Lobo, Classical Quantum Gravity 23,
 	5811 (2006); M. Cataldo, P. Meza, and P. Minning, Phys.
 	Rev. D 83, 044050 (2011).
 	
 	\bibitem{Mehdizadeh:2015dta} 
  M.~R.~Mehdizadeh, M.~Kord Zangeneh and F.~S.~N.~Lobo,
  Phys.\ Rev.\ D {\bf 92}, no. 4, 044022 (2015)
  [arXiv:1506.03427 [gr-qc]].
  
 	\bibitem{thin1} S. H. Mazharimousavi, M. Halilsoy, and Z. Amirabi, Phys.
 	Rev. D 81, 104002 (2010); Classical Quantum Gravity 28,
 	025004 (2011); M. R. Mehdizadeh, M. K. Zangeneh, and
 	F. S. N. Lobo, Phys. Rev. D 92, 044022 (2015).
 	\bibitem{bran1} A. G. Agnese and M. La Camera, Phys. Rev. D 51, 2011
 	(1995); K. K. Nandi, A. Islam, and J. Evans, Phys. Rev. D
 	55, 2497 (1997); F. S. N. Lobo and M. A. Oliveira, Phys.
 	Rev. D 81, 067501 (2010); S. V. Sushkov and S. M.
 	Kozyrev, Phys. Rev. D 84, 124026 (2011).
 	\bibitem{brn1} E. F. Eiroa and G. F. Aguirre, Eur. Phys. J. C 72, 2240
 	(2012); M. Richarte and C. Simeone, Phys. Rev. D 80,
 	104033 (2009).
 	\bibitem{gaus1} M. R. Mehdizadeh, M. K. Zangeneh, and F. S. N. Lobo,
 	Phys. Rev. D 91, 084004 (2015); M. K. Zangeneh, F. S. N.
 	Lobo, and M. H. Dehghani, Phys. Rev. D 92, 124049
 	(2015).
 	\bibitem{kul1}  V. D. Dzhunushaliev and D. Singleton, Phys. Rev. D 59,
 	064018 (1999); J. P. de Leon, J. Cosmol. Astropart. Phys. 11
 	(2009) 013.
 	\bibitem{scal1}  R. Shaikh and S. Kar, Phys. Rev. D 94, 024011 (2016).
 
 	\bibitem{Bronnikov:2015pha} 
  K.~A.~Bronnikov and A.~M.~Galiakhmetov,
  Grav.\ Cosmol.\  {\bf 21}, no. 4, 283 (2015)
 
 	\bibitem{Bronnikov:2016xvj} 
  K.~A.~Bronnikov and A.~M.~Galiakhmetov,
  Phys.\ Rev.\ D {\bf 94}, no. 12, 124006 (2016)
  
 	\bibitem{ecart1} M. R. Mehdizadeh and A. H. Ziaie, Phys. Rev. D 95,
 	064049 (2017).
 	\bibitem{accel1} C. Deffayet, G. R. Dvali, and G. Gabadadze, Phys. Rev. D65, 044023 (2002); S. M. Carroll, V. Duvvuri, M. Trodden, and M. S. Turner, Phys. Rev. D70, 043528 (2004);  S. Nojiri and S. D. Odintsov, Phys. Rev. D68, 123512 (2003);
 	\bibitem{Starobinsky:1980te} 
  A.~A.~Starobinsky,
  Phys.\ Lett.\ B {\bf 91}, 99 (1980)
 	\bibitem{mod1} V. Faraoni, Phys. Rev. D72, 124005 (2005); A. de la Cruz-Dombriz and A. Dobado, Phys. Rev. D74,
 	087501 (2006); N. J. Poplawski, Phys. Rev. D74, 084032 (2006); T. Clifton and J. D. Barrow, Phys. Rev. D72, 103005
 	(2005); S. Nojiri and S. D. Odintsov, Phys. Rev. D74, 086005
 	(2006); G. Cognola, E. Elizalde, S. Nojiri, S. D. Odintsov,
 	and S. Zerbini, Phys. Rev. D73, 084007 (2006).
 	\bibitem{lobener} T. Harko, F. S. N. Lobo, M. K. Mak, S. V. Sushkov, Phys. Rev. D 87, 067504 (2013); F.S.N. Lobo,
 	AIP Conf. Proc. 1458, 447 (2011); F.S.N. Lobo, M. A. Oliveira, Phys. Rev. D 80, 104012 (2009)
 	\bibitem{deb} N. Furey, A. De Benedictis, Class. Quant. Grav. 22, 313 (2005); A. De Benedictis, D. Horvat, Gen. Rel.
 	Grav. 44, 2711 (2012).
 	\bibitem{rahm1} F. Rahaman, A. Banerjee, M. Jamil, A. K. Yadav, H. Idris, Int. J. Theor. Phys. 53, 1910 (2014); M.
 	Jamil, F. Rahaman, R. Myrzakulov, P.K.F. Kuhfittig, N. Ahmed, U.F. Mondal, J. Korean Phys. Soc.
 	65, 917 (2014).
 	\bibitem{char} S. Bhattacharya, S. Chakraborty, Eur. Phys. J. C 77:558 (2017);S. Bahamonde, M. Jamil, P. Pavlovic, and M. Sossich, Phys. Rev. D 94 044041 (2016):
 	\bibitem{ero2}  E.F. Eiroa and G. Figueroa Aguirre, Eur. Phys. J. C 78, 54 (2018);  E.F. Eiroa and G. Figueroa Aguirre, Eur. Phys. J. C 76, 132 (2016); E.F. Eiroa and G.
 	Figueroa Aguirre, Phys. Rev. D 94, 044016 (2016);
 	\bibitem{ero1} E.F. Eiroa, G. Figueroa Aguirre, and J.M.M. Senovilla, Phys. Rev. D 95, 124021 (2017).
 	\bibitem{aziz1} T. Azizi, Int. J. Theo. Phys. 52, 3486 (2013); N.M. Garcia, F. S. N. Lobo, Phys. Rev. D 82, 104018
 	(2010).
 	\bibitem{pmarko} P. Pavlovic, M. Sossich, Eur.Phys.J. C75 117 (2015).
 	\bibitem{solar} O. Bertolami and J. Paramos, Class. Quant. Grav. 25 (2008) 245017;  O. Bertolami and J. Paramos, Phys. Rev. D 77 (2008) ; T. Multamaki and I. Vilja, Phys. Rev. D73, 024018
 	(2006);G. J. Olmo, Phys. Rev. D75, 023511 (2007);J. A. R. Cembranos, Phys. Rev. D73, 064029 (2006); W. Hu and I. Sawicki, Phys. Rev. D 76, 064004 (2007).
 	\bibitem{saba} Lorenzo Sebastiani, Luciano Vanzo, Sergio Zerbini, Phys.Rev. D 97, 044009 (2018).
 	\bibitem{Aclas} Marco Calz�, Massimiliano Rinaldi, Lorenzo Sebastiani, Eur.Phys.J. C78, 178 (2018).

  \bibitem{Bronnikov:2006pt} 
  K.~A.~Bronnikov and A.~A.~Starobinsky,
  JETP Lett.\  {\bf 85}, 1 (2007)
  [Pisma Zh.\ Eksp.\ Teor.\ Fiz.\  {\bf 85}, 3 (2007)]
 
 \bibitem{Bronnikov:2010tt} 
  K.~A.~Bronnikov, M.~V.~Skvortsova and A.~A.~Starobinsky,
  Grav.\ Cosmol.\  {\bf 16}, 216 (2010)
\bibitem{Lobo:2009ip} 
  F.~S.~N.~Lobo and M.~A.~Oliveira,
 ``Wormhole geometries in f(R) modified theories of gravity,''  Phys.\ Rev.\ D {\bf 80}, 104012 (2009)
  [arXiv:0909.5539 [gr-qc]].
  
\bibitem{Tsujikawa:2007xu} 
  S.~Tsujikawa,
  Phys.\ Rev.\ D {\bf 77}, 023507 (2008)
  [arXiv:0709.1391 [astro-ph]].
  
  \bibitem{DeFelice:2010aj} 
  A.~De Felice and S.~Tsujikawa,
  ``f(R) theories,''
  Living Rev.\ Rel.\  {\bf 13}, 3 (2010)
  [arXiv:1002.4928 [gr-qc]].
  
  \bibitem{luca}
  L. Amendola and S. Tsujikawa 2010
  ``DARK ENERGY Theory and Observations'', Cambridge University Press, 2010.
  
 \bibitem{Starobinsky:2007hu} 
  A.~A.~Starobinsky,
  ``Disappearing cosmological constant in f(R) gravity,''
  JETP Lett.\  {\bf 86}, 157 (2007)
  [arXiv:0706.2041 [astro-ph]].
  
  \bibitem{Amendola:2006kh} 
  L.~Amendola, D.~Polarski and S.~Tsujikawa,
  ``Are f(R) dark energy models cosmologically viable ?,''
  Phys.\ Rev.\ Lett.\  {\bf 98}, 131302 (2007)
  [astro-ph/0603703].
  
   \bibitem{Hu:2007nk} 
  W.~Hu and I.~Sawicki,
 ``Models of f(R) Cosmic Acceleration that Evade Solar-System Tests,''
  Phys.\ Rev.\ D {\bf 76}, 064004 (2007)
  [arXiv:0705.1158 [astro-ph]].
   
  \bibitem{n-o} 
  S.~Nojiri and S.~D.~Odintsov,
  ``Modified f(R) gravity unifying R**m inflation with Lambda CDM epoch,''
  Phys.\ Rev.\ D {\bf 77}, 026007 (2008)
  [arXiv:0710.1738 [hep-th]].
  
  \bibitem{agp} 
  L.~Amendola, R.~Gannouji, D.~Polarski and S.~Tsujikawa,
 ``Conditions for the cosmological viability of f(R) dark energy models,''
  Phys.\ Rev.\ D {\bf 75}, 083504 (2007)
  [gr-qc/0612180].
  
  \bibitem{Cognola:2007zu} 
  G.~Cognola, E.~Elizalde, S.~Nojiri, S.~D.~Odintsov, L.~Sebastiani and S.~Zerbini,
 ``A Class of viable modified f(R) gravities describing inflation and the onset of accelerated expansion,''
  Phys.\ Rev.\ D {\bf 77}, 046009 (2008)
  [arXiv:0712.4017 [hep-th]].
  
  \bibitem{Elizalde:2010ts} 
  E.~Elizalde, S.~Nojiri, S.~D.~Odintsov, L.~Sebastiani and S.~Zerbini,
 ``Non-singular exponential gravity: a simple theory for early- and late-time accelerated expansion,''
  Phys.\ Rev.\ D {\bf 83}, 086006 (2011)
  [arXiv:1012.2280 [hep-th]].
  
  \end{thebibliography}
 \end{document}